\documentclass[12pt,a4paper,final]{iopart} 
\newcommand{\Jpsi}{J/\psi}
\newcommand{\pT}{p_{T}}
\newcommand{\sNN}{\sqrt{s_{_{NN}}}}
\newcommand{\raa}{R_{AA}}
\newcommand{\ccbar}{\rm{c\overline{c}}}
\newcommand{\bbbar}{\rm{b\overline{b}}}
\newcommand{\npart}{N_{\rm part}}

\usepackage{verbatim}
\usepackage{graphicx}
\usepackage[usenames,dvipsnames,svgnames,table]{xcolor}
\usepackage[breaklinks=true,colorlinks=true,linkcolor=blue,urlcolor=blue,citecolor=blue]{hyperref}
\usepackage{rotating}
\usepackage{graphicx}
\usepackage{dcolumn}
\usepackage{bm}
\usepackage{epsfig}
\usepackage{hyperref}
\usepackage{ulem}
\usepackage{appendix}
\usepackage{iopams}
\usepackage{subfig}
\expandafter\let\csname equation*\endcsname\relax
\expandafter\let\csname endequation*\endcsname\relax
\usepackage{amsmath}

\begin{document}

\title[]{Suppression of quarkonia in PbPb collisions at $\sqrt{s_{NN}}$ =  5.02 TeV}
\author{Vineet Kumar}
\address{Nuclear Physics Division, Bhabha Atomic Research Center, Mumbai, India}
\address{Department of Physics, University Of Illinois, Chicago, USA}
\ead{Vineet.Kumar@cern.ch}
\author{Prashant Shukla}
\address{Nuclear Physics Division, Bhabha Atomic Research Center, Mumbai, India}
\address{Homi Bhabha National Institute, Anushakti Nagar, Mumbai, India}
\author{Abhijit Bhattacharyya}
\address{Department of Physics, University of Calcutta, 92, A. P. C. Road Kolkata-700009, India}

\date{\today}

\begin{abstract}
  
  We study different processes responsible for the modification of quarkonia yields
in the medium produced in PbPb collisions at $\sqrt{s_{NN}}$ =  5.02 TeV.
The quarkonia and heavy flavour cross sections are estimated using the measurements in
pp collisions at LHC energies and shadowing corrections are obtained using the EPS09
parameterizations. A kinetic model is used which incorporates quarkonia suppression
inside the Quark Gluon Plasma (QGP), suppression due to hadronic comovers, and regeneration
from recombination of heavy quark pairs. The quarkonia dissociation cross section due to gluon 
collisions, including both color-electric dipole and color-magnetic dipole transitions,
has been employed. The regeneration rate is obtained using the principle of 
detailed balance. The effect of these processes on the nuclear modification
factors for both $\Jpsi$ and $\Upsilon$ in different ranges of transverse momentum
$\pT$ and rapidity has been studied for PbPb collisions at $\sNN =$ 5.02 TeV. The calculations
are compared with the available results from LHC experiments. Both the suppression
and regeneration due to a QGP are effective in the low and intermediate $\pT$ range.
The large observed suppression of $\Jpsi$ at
$\pT~>$ 10 GeV/$c$ is larger than the suppression expected due to gluon dissociation.

\end{abstract}


\maketitle

\section{Introduction}

Quantum chromodynamics (QCD) predicts that strongly interacting matter undergoes a phase
transition to quark-gluon plasma (QGP), a state in which quarks and gluons move much beyond the
size of hadrons. Heavy-ion collisions at relativistic energies are performed to create and
quantify the properties of QGP~\cite{Busza:2018rrf,Shuryak:2017aol}. Charmonia and bottomonia,
which are bound states of charm-anticharm (c$\bar{\rm c}$) or bottom-antibottom (b$\bar{\rm b}$) quarks,
respectively, are among the most sensitive probes of the characteristics of the QGP~\cite{Brambilla:2010cs}.
These bound states of heavy quarks are formed early in the heavy ion collisions and their
yields are expected to be suppressed in the medium as compared to their yields in pp
collisions ~\cite{Matsui:1986dk,Hashimoto:1986nn}. 
 There have been a large number of studies on this phenomenon both theoretically and 
 experimentally~\cite{Brambilla:2010cs,Schukraft:2013wba,Andronic:2015wma} 
 enriching our understanding on quarkonia as probes of QGP. 
 The J/$\psi$ meson was measured at the SPS, in PbPb and In-In interactions
at the centre-of-mass energy per nucleon pair $\sNN$ = 17.2 GeV~\cite{Alessandro:2004ap,Arnaldi:2007zz},
at RHIC in AuAu interactions at 
$\sNN$ = 200 GeV~\cite{Adare:2011yf,Abelev:2009qaa,Tang:2011kr}, and finally at the LHC,
in PbPb collisions at $\sNN$ = 2.76 and
5.02 TeV~\cite{Aad:2010aa,Chatrchyan:2012np,Khachatryan:2016ypw,Sirunyan:2017isk,ATLAS:2016qpn,Abelev:2013ila,Adam:2016rdg,Acharya:2017tgv}.
The suppression of $\Jpsi$ observed at SPS was termed as 'anomalous' $\Jpsi$
suppression. It was even considered to be a hint of QGP~\cite{Alessandro:2004ap,Arnaldi:2007zz} formation.
Early theoretical calculations predicted $\Jpsi$ suppression due to colour screening
in a deconfined medium which become stronger as the QGP 
temperature increases~\cite{Matsui:1986dk,Digal:2001ue} but the RHIC measurements
at $\sNN$ = 200 GeV showed almost the same amount of suppression, contrary to
expectation~\cite{Brambilla:2010cs,Adare:2011yf}. These observations hint a scenario where,
at higher collision energies, the expected larger suppression is compensated by
$\Jpsi$ regeneration via recombination of two independently produced 
charm quarks~\cite{Andronic:2003zv,Thews:2000rj,Du:2017qkv}. 

The LHC collected first PbPb collision data at $\sNN =$ 2.76 TeV at the end of 2010.
The ATLAS was the first detector to measure 
the ratio of $\Jpsi$ meson with Z$^{0}$ boson hinting a centrality-dependent suppression
of the yield of $\Jpsi$ mesons~\cite{Aad:2010aa}.
 The $\Jpsi$ measurements at high transverse momentum ($p_T>6.5$ GeV/$c$) in PbPb collisions at $\sNN$=2.76 TeV and at
$\sNN$=5.02 TeV were carried out by the CMS experiment~\cite{Chatrchyan:2012np,Khachatryan:2016ypw,Sirunyan:2017isk}.
The nuclear modification factor $R_{AA}$ of these high $p_T$ prompt $\Jpsi$ decreases
with increasing centrality. The $R_{AA}$ shows a slow increase as a function of
$p_T$ after 15 GeV$/c$ and then saturates between a value 0.4 and 0.5 showing that the $\Jpsi$
remains suppressed, even at very high $p_T$, up to $\sim$ 50 GeV/$c$. The ATLAS experiment
also measured $R_{AA}$ of $\Jpsi$ at $\sNN$=5.02 TeV for $\Jpsi$ meson having
transverse momentum above 9.0 GeV$/c$~\cite{ATLAS:2016qpn}. The amount 
of suppression of $\Jpsi$ mesons observed by ATLAS is similar to the suppression
observed by CMS experiment. By comparing these measurements with the STAR
results~\cite{Tang:2011kr} at RHIC, it follows that at high $\pT$ the suppression of 
$\Jpsi$ increases with collision energy.

  The ALICE experiment measured the nuclear modification factor of $\Jpsi$ mesons 
in the forward rapidity (2.5$<y<$4.0) range at $\sNN$=2.76 and $\sNN$=5.02 
TeV~\cite{Abelev:2013ila,Adam:2016rdg} with $\Jpsi$ transverse momentum starting from
0.3 GeV/$c$.
The ALICE results show that nuclear modification factor of J/$\psi$ at 
low $\pT$ ($\pT<$ 12 GeV/$c$) has almost no collision centrality dependence 
except in the most peripheral region where it reaches almost unity.
The $R_{AA}$ of $\Jpsi$ mesons decreases substantially as a function of $\pT$ 
in the ALICE experiment~\cite{Abelev:2013ila,Adam:2016rdg}. The ALICE measurements also
give a hint for an increase of $R_{AA}$ between $\sNN$ = 2.76 and 5.02 TeV in the intermediate
$\pT$ region, 2 $< \pT < $ 6 GeV/$c$. A comparison of 
the ALICE and PHENIX measurements reveal that at LHC, $\Jpsi$ mesons with low $p_T$ are less
suppressed as compared to RHIC. In general, the results at LHC experiments when 
compared to the RHIC measurements indicate that the data on $\Jpsi$ production support a 
picture where both suppression and regeneration take place in the QGP, 
the two mechanisms being dominant at high and low $\pT$, respectively~\cite{P.ShuklaforCMS:2014vna,Kumar:2014kfa}.

In addition to $\Jpsi$ mesons, the bottomonia states ($\Upsilon$(nS)) are also measured at
the LHC with very good statistical
precision~\cite{Chatrchyan:2011pe,Chatrchyan:2012lxa,Abelev:2014nua,Khachatryan:2016xxp}.
The CMS measurements at $\sNN =$2.76 TeV~\cite{Chatrchyan:2011pe,Chatrchyan:2012lxa} reveal
a clear proof of sequential suppression :  $\Upsilon$(2S) and $\Upsilon$(3S) are 
more suppressed relative to the ground state $\Upsilon$(1S).   The individual $\Upsilon$ states are also found to be suppressed in
the PbPb collisions relative to the production in the pp collisions. The $\Upsilon$ nuclear
modification factor, $R_{AA}$, shows a strong dependence on collision centrality but has
weak dependence on $\Upsilon$ meson $\pT$ and rapidity~\cite{Khachatryan:2016xxp}.
The forward rapidity ($2.5 \leq y^{\Upsilon} \leq 4.0$) measurement of the $\Upsilon$ suppression at 
ALICE~\cite{Abelev:2014nua} is found to be consistent with the midrapidity ($|y^{\Upsilon}|\,\leq 2.4$)
measurement of the $\Upsilon$ suppression at the CMS. 
The CMS and ALICE collaborations have carried out the $R_{AA}$ measurement of $\Upsilon$
at $\sNN =$ 5.02 TeV with the Run II LHC PbPb
collisions~\cite{Sirunyan:2017lzi,Sirunyan:2018nsz,ALICE:Y5TeV}.
The CMS experiment measured slightly more amount of $\Upsilon$ suppression at
$\sNN =$ 5.02 TeV~\cite{Sirunyan:2017lzi,Sirunyan:2018nsz} than the suppression at
$\sNN =$ 2.76 TeV~\cite{Khachatryan:2016xxp} while the ALICE experiment observed less
suppression at $\sNN =$ 5.02 TeV than that at $\sNN =$ 2.76 TeV 
in the most central PbPb collisions~\cite{Abelev:2014nua,ALICE:Y5TeV}. 

 The field of quarkonia has attracted a large amount of theoretical activities
since the first prediction of $\Jpsi$ suppression in heavy 
ion collisions~\cite{Matsui:1986dk,Hashimoto:1986nn}. 
The color screening
of q$\overline{\rm q}$ potential inside the QGP~\cite{Karsch:1987pv,Abdulsalam:2012bw}
was the first approach to study the suppression of quarkonia
in heavy ion collisions. 
 In a complementary way to this static approach, $\Jpsi$ suppression can also be understood
as a result of dynamical interactions with the surrounding gluons~\cite{Bhanot:1979vb,Chen:2017jje,Kumar:2014kfa}.
One of the ways to calculate the regeneration is to use the principle of detailed balance~\cite{Thews:2000rj}.  
There are other effects also, namely shadowing and comover interaction~\cite{Vogt:2015uba,Ferreiro:2014bia}.
Shadowing arises as the parton distribution functions are modified inside the nucleus. 
A comprehensive framework to explain the experimental data from LHC considering shadowing and 
comover~\cite{Du:2017qkv,Rapp:2017chc} and viscous hydrodynamics~\cite{Krouppa:2015yoa,Krouppa:2017jlg}
has recently been formulated. 

 Some of us have studied the modification of quarkonia yields due to various
processes in AA collisions in a previous work~\cite{Kumar:2014kfa}.
In that work, the gluon dissociation cross section has
been adopted from the calculations of Bhanot and Peskin~\cite{Bhanot:1979vb}
where the gluon dissociation rate has been estimated from operator product expansion
in the Coulomb approximation. Recently, Chen and He have revisited the gluon dissociation
using QCD multipole expansion for various quarkonia in QGP~\cite{Chen:2017jje}.
They reproduced the result of Bhanot and Peskin as the colour-electric 
dipole (E1) transition. They have also calculated the colour magnetic
dipole (M1) transition and found its contribution to be significant
at low energies.
 In this paper, we calculate $\Jpsi$ and $\Upsilon$ evolution
in a kinetic model which includes dissociation by thermal gluons
(both E1 and M1 transitions), modification 
of their yields due to shadowing and due to collisions with comovers.
Regeneration by near thermal heavy quark pairs is also considered in the calculations.
We obtain the nuclear modification factor of quarkonia as a function of the
transverse momentum and centrality of the collision and compare it to the latest
experimental data from LHC at $\sNN =$ 5.02 TeV.

\section{Theoretical formulation}

The theoretical formulation towards the formation and dissociation of quarkonia have already been discussed 
in Ref.~\cite{Kumar:2014kfa}. We use the same formulation here and hence discuss only some important points.

\subsection{Formation and Dissociation}

In the kinetic approach \cite{Thews:2000rj}, the evolution of the quarkonia 
population $N_{Q}$ with the proper time, $\tau$, is given by the kinetic equation

\begin{equation}\label{eqkin}
{dN_{Q} \over d\tau}  =  - \lambda_D  \rho_g N_{Q} + \lambda_F {N_{q \bar{q}}^{2} \over V(\tau)}
\end{equation}
In the above equation $V(\tau)$ is the spatial volume of the QGP  and $N_{q \bar{q}}$ is
the number of initial heavy quark pairs produced in a event as a function of the centrality
defined by the number of participants $N_{\rm part}$. The $\lambda_{D}$ is the dissociation
rate and the $\lambda_{F}$ is the formation rate. Here $\rho_g$ is the density of thermal gluons.

\begin{figure}
\includegraphics[width=0.49\textwidth]{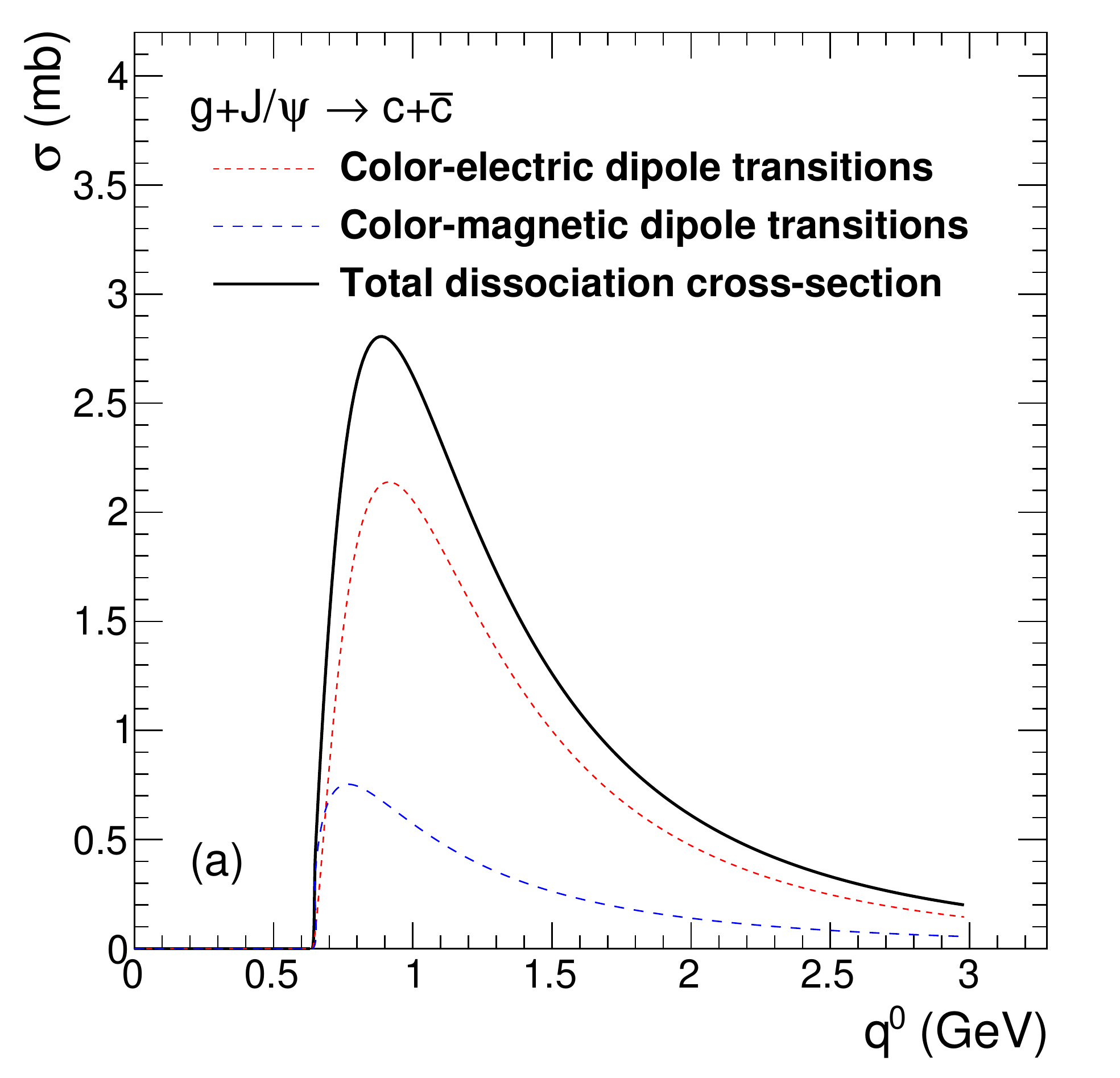}
\includegraphics[width=0.49\textwidth]{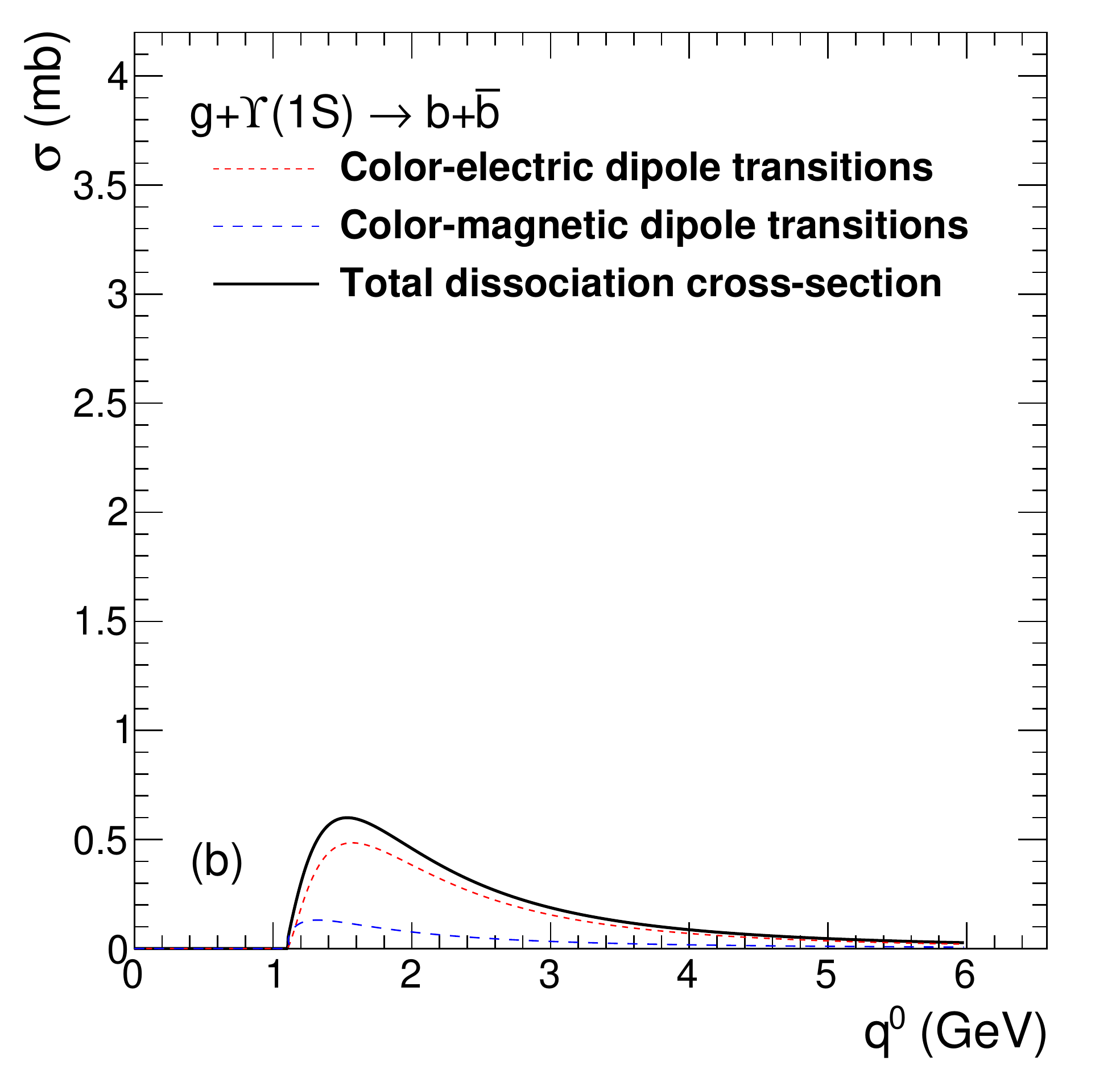}
\caption{(Color online) Gluon dissociation cross section of J/$\psi$ (a) and $\Upsilon$(1S) (b) as a function of gluon energy
    ($q^{0}$) in quarkonia rest frame. The total dissociation cross-section is sum of both color-electric and chromo-magnetic dipole 
    transitions.}
\label{fig:SigmaDQ0}
\end{figure}

The gluon dissociation cross section of quarkonia was calculated by
Bhanot and Peskin using operator product expansion method in the color dipole
approximation~\cite{Bhanot:1979vb}. Recently, Chen and He have revisited the
gluon dissociation using QCD multipole expansion method~\cite{Chen:2017jje}.
They reproduced the result of Ref.~\cite{Bhanot:1979vb} as the color-electric dipole (E1)
transition and also calculated the color magnetic dipole (CMD-M1) transition.
The E1 transition cross-section of gluon dissociation as a function of gluon energy, $q^0$,
in the quarkonium rest frame is~\cite{Bhanot:1979vb}
\begin{equation}
\sigma^{E1}_{D}(q^{0}) = {8\pi \over 3} \, {16^2 \over 3^2} {a_0 \over m_q}  \frac{(q^0/\epsilon_0 - 1)^{3/2}} {(q^0/\epsilon_0)^5}
\end{equation}
 where $\epsilon_0$ is the quarkonia binding energy and $m_q$ is the charm/bottom quark mass 
and $a_0=1/\sqrt{m_q\epsilon_0}$. Using the same procedure, the M1 transition cross-section of gluon 
dissociation can be calculated as~\cite{Chen:2017jje}
\begin{equation}
\sigma^{M1}_{D}(q^{0}) = {8\pi \over 3} \, {16 \over 3 } {a_0 \over m_q^{2}}  \frac{\epsilon_0(q^0/\epsilon_0 - 1)^{3/2}} {(q^0/\epsilon_0)^3}
\end{equation}
The total dissociation cross section is given as 
\begin{equation}
\sigma_D(q^0) = \sigma^{E1}_D(q^0) + \sigma^{M1}_D(q^0).
\end{equation}
The values of $\epsilon_0$ are taken as 0.64 GeV and 1.10 GeV for the ground states, $\Jpsi$ and $\Upsilon$(1S),
respectively~\cite{Karsch:1987pv}. For the excited states of bottomonia, the dissociation cross sections are
used from Ref.~\cite{Chen:2017jje,Oh:2001rm}.

\begin{figure}
\includegraphics[width=0.49\textwidth]{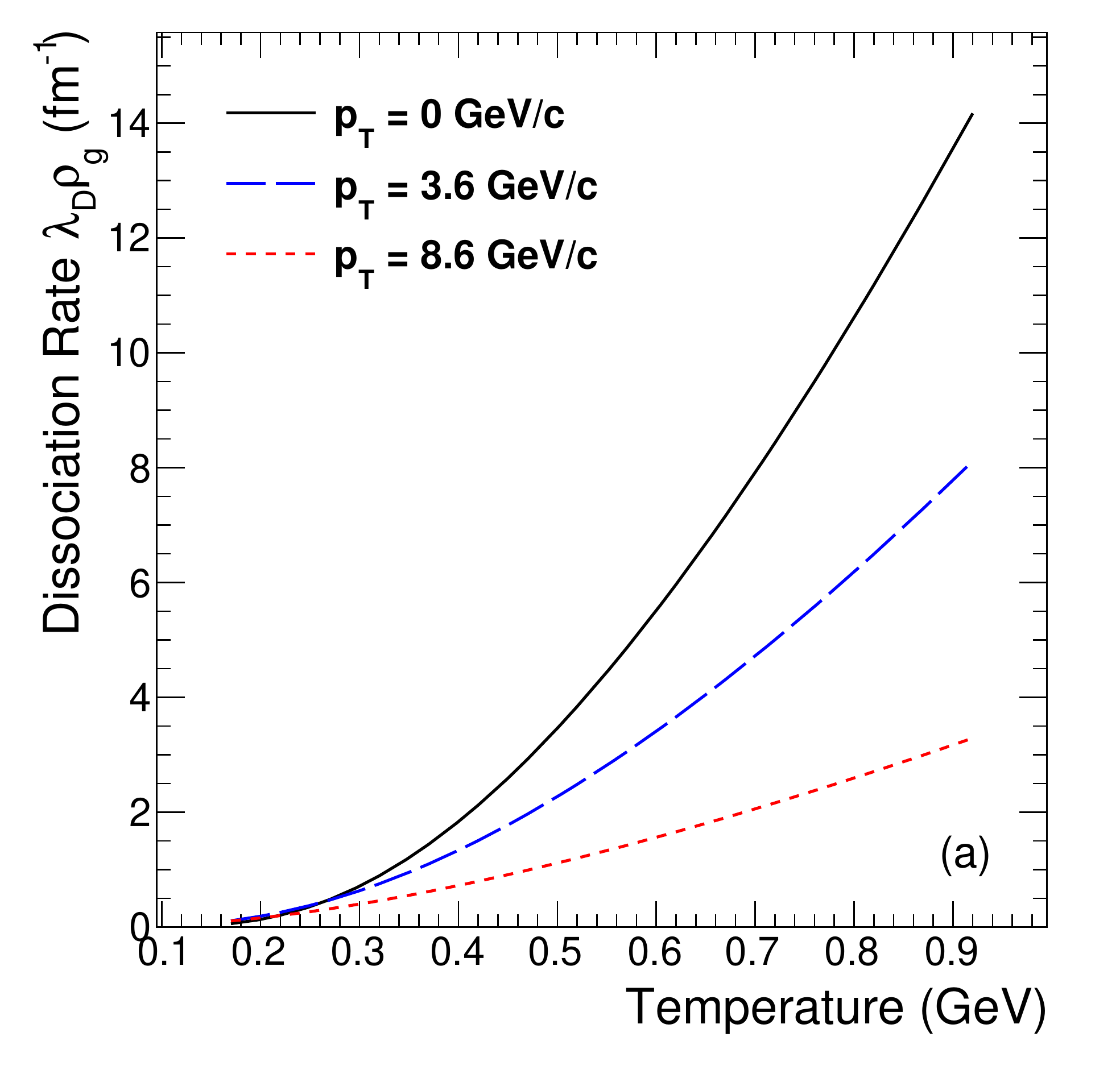}
\includegraphics[width=0.49\textwidth]{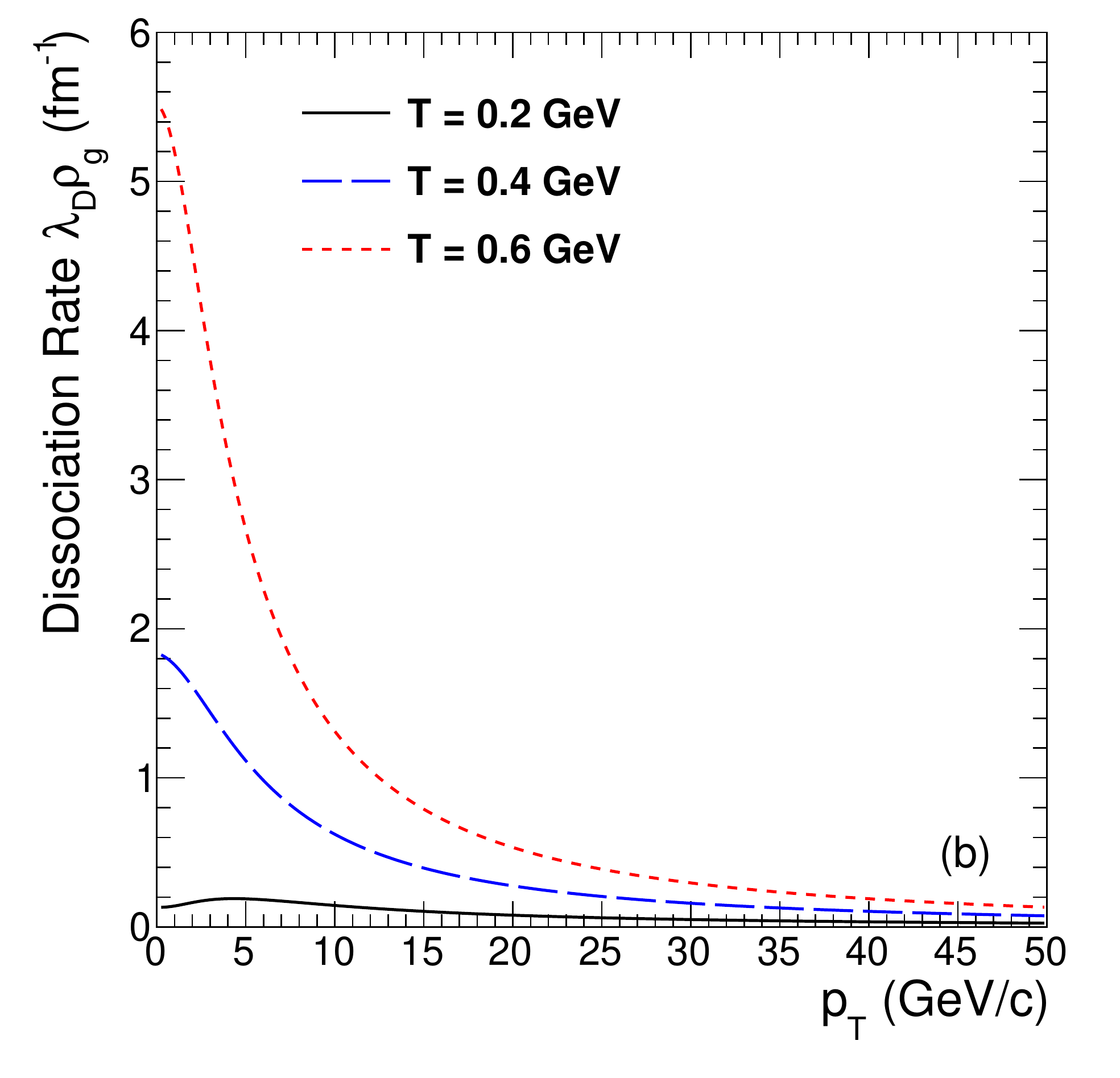}
\caption{(Color online) Gluon dissociation rate of $\Jpsi$ as a function of (a) temperature and  
(b) $\Jpsi$ transverse momentum.}
\label{fig:DRateVsTempAndPt}
\end{figure}

Figure~\ref{fig:SigmaDQ0} shows the gluon dissociation cross sections
of $\Jpsi$ and $\Upsilon$(1S) as a function of gluon energy
for both color-electric and color-magnetic dipole transitions.
The color-magnetic dipole transition cross-section has similar shape as that of
electric cross-section and gives significant contribution 
in low and intermediate gluon energies. Total dissociation cross section increases
with gluon energy and reaches a maximum around 0.9 GeV for
J/$\psi$ and around 1.5 GeV for $\Upsilon$(1S). At higher gluon energies, the
interaction probability decreases. 
 The dissociation rate is calculated as a function of quarkonium momentum 
by integrating the dissociation cross section over thermal gluon momentum 
distribution $f_{g}(p_g)$ as 
\begin{eqnarray}
\lambda_{D} \rho_{g}  & = & \langle \sigma v_{\rm rel} \rangle \,\rho_{g}  = \frac{g_g}{(2\pi)^{3}} \int d^{3}p_g \, f_{g}(p_g)  \, \sigma_{D}(s) v_{\rm rel}(s)  \nonumber \\ 
                   & = & \frac{g_g}{(2\pi)^{3}} \int dp_g 2\pi p_g^{2} f_{g}(p_g) \int d\,{\rm cos\theta}\,\sigma_{D}(s)\,v_{\rm rel}(s),
\label{Eqn:Diss}
\end{eqnarray}
where $\sigma_{D}(s) = \sigma_{D}(q^0(s))$ in terms of the square of the center of
mass energy $s$ of the quarkonium-gluon system given by
$s=M_{Q}^{2} + 2  p_g \, \sqrt{M_{Q}^2 + p^2} - 2  p_g \, p \, {\rm cos\theta}$.
Here $M_{Q}$ is the mass and $p$ is the momentum of quarkonium and $\theta$ is the angle
between the quarkonium and the gluon.
  The variables $q^0$ and $s$ are related by $q^{0} = (s-M_{Q}^{2})/(2\,M_{Q})$. 
$v_{\rm rel}$ is the relative velocity between the quarkonium and the gluon~\cite{Kumar:2014kfa}.
  The formula in Eq.~\ref{Eqn:Diss} is assumed
for the most central collisions. We multiplied by a system size dependent factor
($\sqrt{N_{\rm part}/2A}$) to get the dissociation rate for other centralities.
The $\Jpsi$ gluon dissociation rates as a function of temperature $T$ are shown in 
Fig.~\ref{fig:DRateVsTempAndPt}(a) and as a function of $p_T$ in Fig.~\ref{fig:DRateVsTempAndPt}(b).
The dissociation rate increases with temperature as the gluon density increases. 
Also, it is maximum when the quarkonium is at rest and then decreases with $p_T$.

We can calculate the formation cross section from the dissociation cross section
using principle of detailed balance~\cite{Thews:2000rj,Thews:2005vj} as follows
\begin{equation}
\sigma_{F} = \frac{48}{36}\,\sigma_{D}(q^0)\frac{(s-M_{Q}^2)^{2}}{s(s-4m_q^{2})}.
\end{equation}
The formation rate of quarkonium at momentum {\bf p} can be written as
\begin{equation}
\frac{d\lambda_{F}}{d{\rm\bf p}} = \int \,d^{3}p_1 \,d^{3}p_2 \,\sigma_{F}(s)\, v_{\rm rel}(s)\,f_{q}(p_1)\, f_{\bar{q}} (p_2)\,\delta({\rm\bf p}-( {\rm\bf p_1} + {\rm\bf p_2} )).
\end{equation}
The variable $s$ is the square of center of mass energy between the two heavy quarks with
energy-momenta ($E_1$, ${\rm\bf p_1}$) and ($E_2$, ${\rm\bf p_2}$) with $v_{\rm rel}$ as their
relative velocity.

\begin{figure}
\begin{center}
  \includegraphics[width=0.60\textwidth]{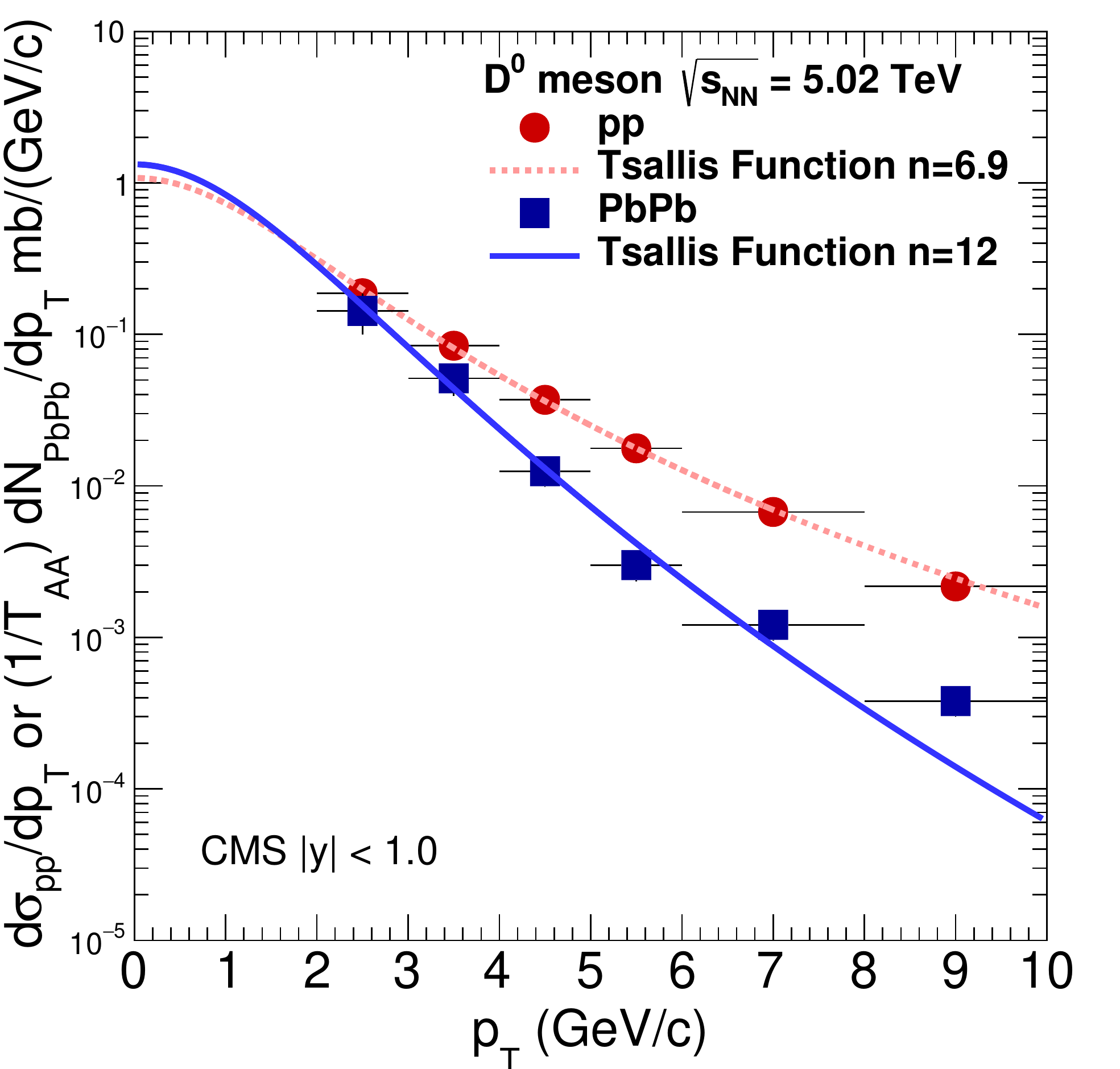}
  \caption{The transverse momentum spectra of D mesons in pp and PbPb collisions
    at $\sqrt{s_{\rm NN}}=5.02$ TeV measured by CMS experiment~\cite{Sirunyan:2017xss}.
    The spectra are fitted by Tsallis function with $n=6.9$ for pp and $n=12$ for PbPb collisions.
}
\label{fig:Figure3_Tsallis}
\end{center}
\end{figure}

The functions $f_{q/\bar{q}}(p)$ are taken as normalized near-thermal distribution functions
of $q/\bar{q}$. These distributions can be described by the Tsallis
function as follows 
\begin{equation}
f_{q} (p,T) = A_{n}\,\left( 1+\frac{ \sqrt{p^2+m_q^2} }{n \,T} \right)^{-n}.
\end{equation}
Here $A_n$ is the normalization factor and $n=12$ is obtained by fitting the
transverse momentum spectra of D mesons measured by CMS experiment~\cite{Sirunyan:2017xss}.
Figure~\ref{fig:Figure3_Tsallis} shows the
transverse momentum spectra of D mesons in pp and PbPb collisions
at $\sqrt{s_{\rm NN}}=5.02$ TeV measured by CMS experiment. The spectra are fitted
by Tsallis function with $n=6.9$ for pp and $n=12$ for PbPb collisions.

\begin{figure}
\includegraphics[width=0.49\textwidth]{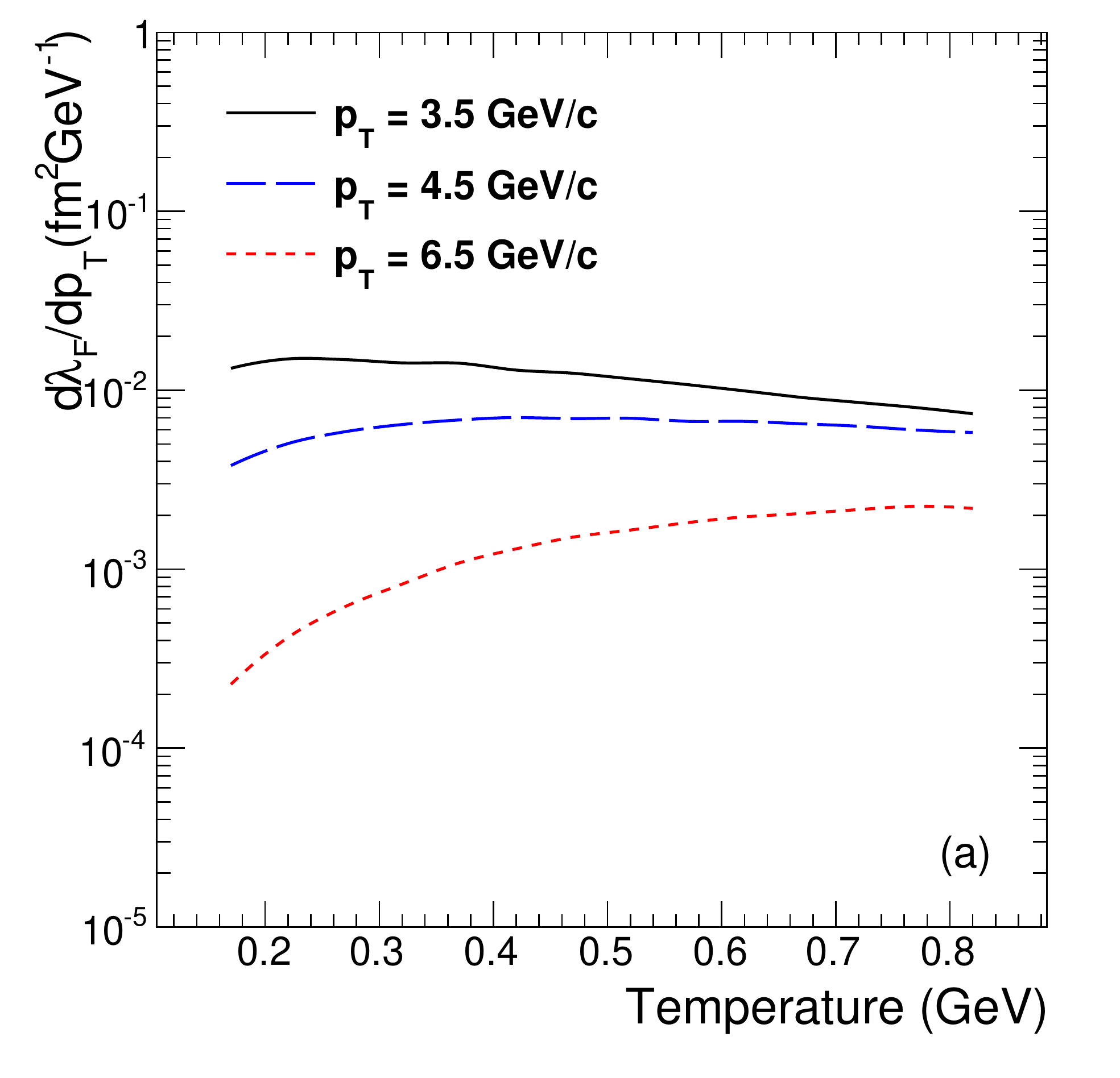}
\includegraphics[width=0.49\textwidth]{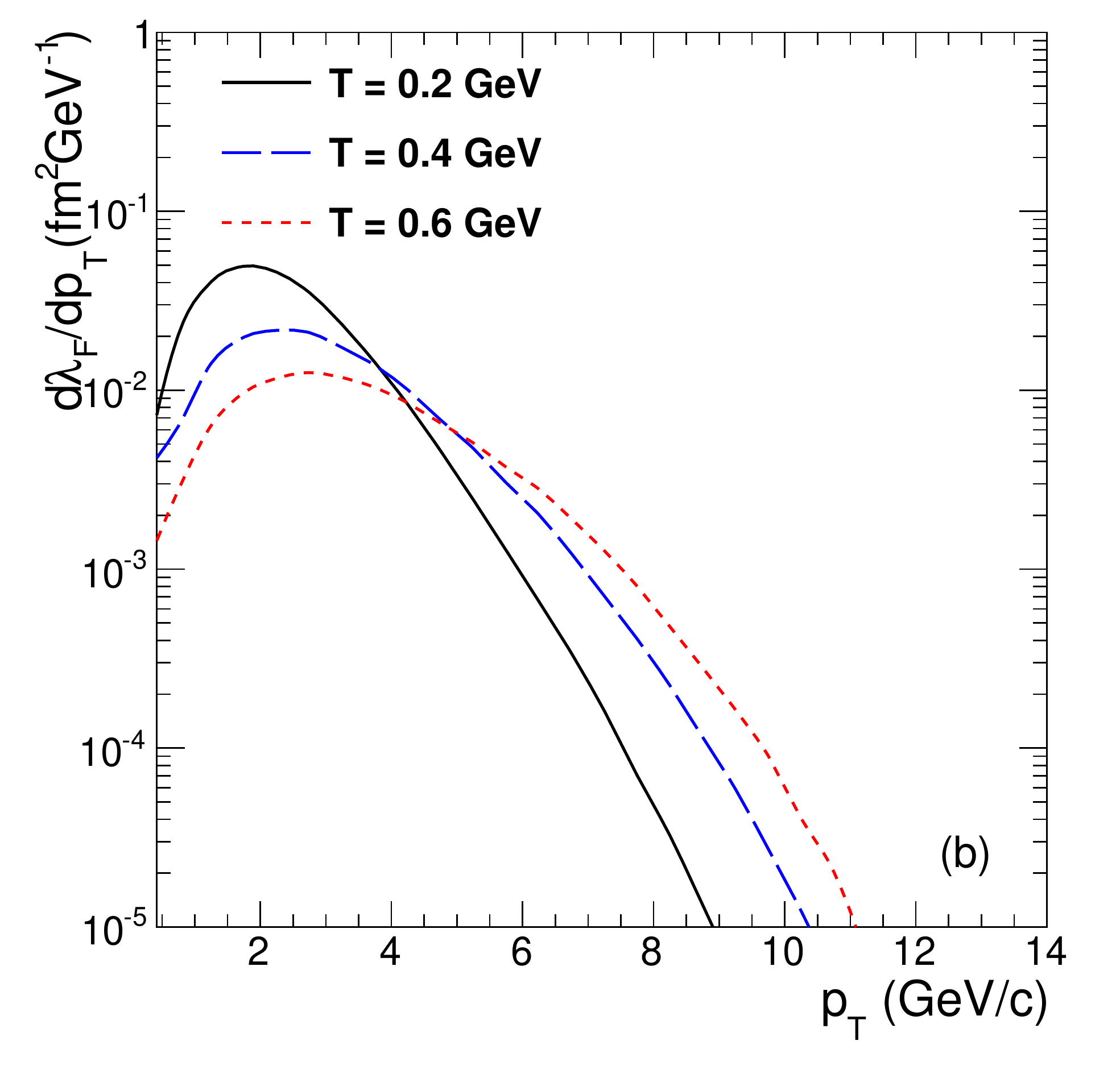}
\caption{(Color online) Formation rate of  $\Jpsi$ as a function of (a) temperature and 
(b) $\Jpsi$ transverse momentum.}
\label{fig:ForRateVsTempAndPt}
\end{figure}

Figure~\ref{fig:ForRateVsTempAndPt}(a) shows the behaviour of the formation rate as a function 
of temperature at different values of $\Jpsi$ meson $\pT$.
Figure~\ref{fig:ForRateVsTempAndPt}(b) shows the same as a function
of $\Jpsi$ meson $p_T$ at different temperatures.
It shows that the $\Jpsi$ generated from uncorrelated heavy quark pairs has 
softer $p_{T}$ distribution than that of $\Jpsi$'s coming from the initial hard scatterings.
Thus the effect of recombination will be important at low and intermediate $p_T$.

\subsection{Nuclear modification}

The nuclear modification factor ($R_{AA}$) as a function of $p_T$ can be obtained as 

\begin{equation}
  R_{AA}(p_T)= \frac{ \Sigma_{\rm Centrality} \, N_Q(p_T,N_{\rm part})}
                    { \Sigma_{\rm Centrality} \, N_{\rm coll} \, N_{Q}^{pp}(p_T)}.
\label{raa}
\end{equation}
The sum over the events is performed over the measured centrality range in the
experiment and N$_{\rm coll}$ is the number of binary collisions for the centrality bins
used in experiments.
 The $\raa$ as a function of collision centrality is obtained as

\begin{equation}
  R_{AA}(N_{\rm part}) = \frac{\int_{p_{T\,\rm cut}} N_{Q}(p_T, \npart) \, p_T dp_T}
                  {\int_{p_{T\,\rm cut}} N_{\rm coll} \,\, N^{\rm pp}_{Q}(p_T) \, p_T dp_T} 
\label{raa2}
\end{equation}
Here $p_{T~{\rm cut}}$ defines the $p_T$ range for acceptance of the experiment.
{\color{black}The shape of quarkonia $p_{T}$ distribution in pp collisions is obtained from
  PYTHIA 6.424~\cite{Sjostrand:2006za}. The PYTHIA generator gives good description of quarkonia data
  at LHC energy~\cite{Khachatryan:2010yr}. The number of Q$\bar{Q}$ pairs in different centrality classes
  are obtained by the N$_{\rm coll}$ scaling. The N$_{\rm coll}$ values in different centrality classes
  are calculated using the Glauber model~\cite{Loizides:2017ack}.
}

The evolution of the system for each centrality range is governed by an isentropic
cylindrical expansion ($s(T)\,V(\tau)= s(T_0)\,V(\tau_0)$) with prescription given
in Ref.~\cite{Kumar:2014kfa}.
The equation of state (EOS) obtained by Lattice QCD and by hadronic resonances is
used~\cite{Huovinen:2009yb}. The radius $R$ for a given centrality
with number of participant $\npart$ is obtained as $R(\npart) = R_{A} \sqrt{\npart/2A}$,
where $R_{A}$ is radius of the nucleus.
The initial entropy density, $s(\tau_0)|_{0-5\%}$, for 0-5\% centrality is 
\begin{eqnarray}
s(\tau_0)|_{0-5\%}  = {a_{m} \over V(\tau_0)|_{0-5\%}}   \left(\frac{dN}{d\eta}\right)_{0-5\%} . 
\label{TempVsMult}
\end{eqnarray}  
Here $a_m$ (= 5) is a constant connecting the total entropy to the final hadron 
multiplicity $dN/d\eta$ obtained from hydrodynamic calculations~\cite{Shuryak:1992wc}.
The initial temperature, $T_0$, in the 0-5$\%$ most central collisions is estimated 
from the total multiplicity in the given rapidity, assuming that the initial time is
$\tau_0 = 0.3$ fm/$c$. The total multiplicity in a given rapidity window is
1.5 times the measured charged particle multiplicity in PbPb collisions at 5.02 TeV.
With the lattice EOS, at midrapidity, with $(dN_{\rm ch}/d\eta)_{0-5\%} = 1943$~\cite{Adam:2015ptt}, 
we find $T_0 = 0.516$ GeV. Likewise, at forward rapidity~\cite{Adam:2016ddh}; $T_0$ is
0.487 GeV. The freeze out temperature is taken to be $T_f=0.140$ GeV.

\subsection{Hadronic comovers}

The suppression of quarkonia caused due to comoving pions is obtained by folding the quarkonium-pion
dissociation cross section $\sigma_{\pi Q}$ over thermal pion distributions \cite{Vogt:1988fj}. 
The  cross section of pion-quarkonia is calculated by convoluting the gluon-quarkonia cross section $\sigma_D$
over the gluon distribution obtained inside the pion~\cite{Arleo:2001mp},
\begin{equation}
\sigma_{\pi Q} (p_{\pi}) = {p_+^2 \over 2(p_\pi^2 - m_\pi^2)} \int_0^1 \, dx \, G(x) \, \sigma_D(xp_+/\sqrt {2}),
\end{equation}
where $p_+ = (p_\pi + \sqrt{p_\pi^2-m_\pi^2})/\sqrt{2}$. The term  $G(x)$ is the gluon distribution inside a pion which 
can be given by the GRV parameterization~\cite{Gluck:1991ey}. 
The comover cross section  is expected  to be small  at LHC energies~\cite{Lourenco:2008sk}.

\subsection{Experimental data to fit}

\begin{table*}[b]
\caption{Total $\ccbar$ production cross-section measured by ALICE experiment in pp collisions at LHC.}
\begin{tabular}{l|l|l|l}
\hline 
\hline
  $\sqrt{s}$(TeV)           &$\sigma^{c\bar{c}}\pm {\rm stat.}\pm {\rm syst.}$(mb)            &Experiment      &Ref.  \\              
\hline
 2.76 TeV                   &$4.8\pm0.8^{+1.0}_{-1.3}$           & ALICE     &\cite{Abelev:2012vra}               \\
 7 TeV                      &$8.5\pm0.5^{+1.0}_{-2.4}$           & ALICE     &\cite{Abelev:2012vra}               \\
 \hline
\hline
\end{tabular}
\label{CCBarCrossExp}
\end{table*}

\begin{table}[t]
\caption[]{Total $\bbbar$ production cross-section measured by ALICE experiment in pp collisions at LHC.}
\label{BBBarCrossExp}
\begin{tabular}{l|l|l|l} 
\hline 
\hline
  $\sqrt{s}$(TeV)           &$\sigma^{\bbbar}\pm {\rm stat.}\pm {\rm syst.}$($\mu$b)            &Experiment      &Ref.  \\              
\hline
 2.76 TeV                   &$130\pm15.1^{+42.1}_{-49.8}$           & ALICE     &\cite{Abelev:2014hla}               \\
 7 TeV                      &$282\pm74^{+58}_{-68}$               & ALICE     &\cite{Abelev:2012gx}               \\
\hline
\hline
\end{tabular}
\end{table}

The total charm and total bottom production cross-sections are measured by different experiments at
LHC~\cite{Abelev:2012vra,Abelev:2014hla,Abelev:2012gx}.
Table~\ref{CCBarCrossExp} and Table~\ref{BBBarCrossExp} show the values of total $\ccbar$ 
and $\bbbar$ production cross-section measured in pp collisions at LHC.
We use these values to estimate the 
total heavy quark production cross section at $\sNN =$ 5.02 TeV.
The quarkonium production cross sections are calculated from the measured heavy 
quark production cross-section using the energy independent factors (0.00526 for J/$\psi$ and 0.002 for $\Upsilon$)
obtained from the color evaporation model~\cite{Kumar:2014kfa,Nelson:2012bc,Vogt:2012vr}.
The cold nuclear matter (CNM) effects i.e. the modifications of the parton distribution
functions (nPDF) in PbPb collisions is calculated using the central EPS09 NLO
parameter set~\cite{Eskola:2009uj}.
The uncertainty in cold matter effects is obtained by adding the EPS09 NLO uncertainties in quadrature.

The production cross sections for heavy flavor and quarkonia at $\sNN =$ 5.02 
TeV are given in Table~\ref{Tab:NLOcross}.  The yields in a minimum bias 
PbPb event is obtained from the per nucleon cross
section, $\sigma_{\rm PbPb}$ as 
\begin{eqnarray}
  N = \frac{A^2 \sigma_{\rm PbPb}}{\sigma_{\rm PbPb}^{\rm tot}} \, \, .
\end{eqnarray}
At 5.02 TeV, the total PbPb cross section, $\sigma_{\rm PbPb}^{\rm tot}$, 
is 7.7 b~\cite{Loizides:2017ack}.





\begin{table}
\caption[]{ Heavy quark and quarkonia production  cross sections per nucleon pair at
  $\sqrt{s_{_{_{NN}}}}= 5.02$ TeV. The quantity $N^{\rm PbPb}$ gives the initial number of
  heavy quark pair/quarkonia in one PbPb event.}
\label{Tab:NLOcross}
\begin{tabular}{l|l|l|l|l} 
\hline 
\hline
             &$ c \overline c$            &J/$\psi$                      & $ b \overline b$                    & $\Upsilon$   \\              
\hline
$\sigma_{pp}$  &$6.754^{+1.195}_{-2.015}$ mb     & $35.32^{+6.247}_{-10.537}~\mu$b     & $210.30^{+70.769}_{-77.640}~\mu$b            & $0.4206^{+0.142}_{-0.155}~\mu$b  \\

&&&&\\
$\sigma_{\rm PbPb}$ &$4.669^{+0.826}_{-1.393}$ mb    &$24.56^{+4.344}_{-7.327}~\mu$b    & $179.30^{+60.337}_{-66.195}~\mu$b             &$0.3586^{+0.121}_{-0.132}~\mu$b  \\
&&&&\\

$N^{\rm PbPb}$     &$26.23^{+4.639}_{-7.825}$       & $0.1381^{+0.024}_{-0.041}$         & $1.007^{+0.339}_{-0.372}$                  & $0.0020^{+0.0007}_{-0.0007}$       \\

\hline
\hline
\end{tabular}
\end{table}

\section{Results and discussions}

\begin{figure}
\begin{minipage}{1.0\linewidth}
\centering
{\includegraphics[width=0.49\textwidth]{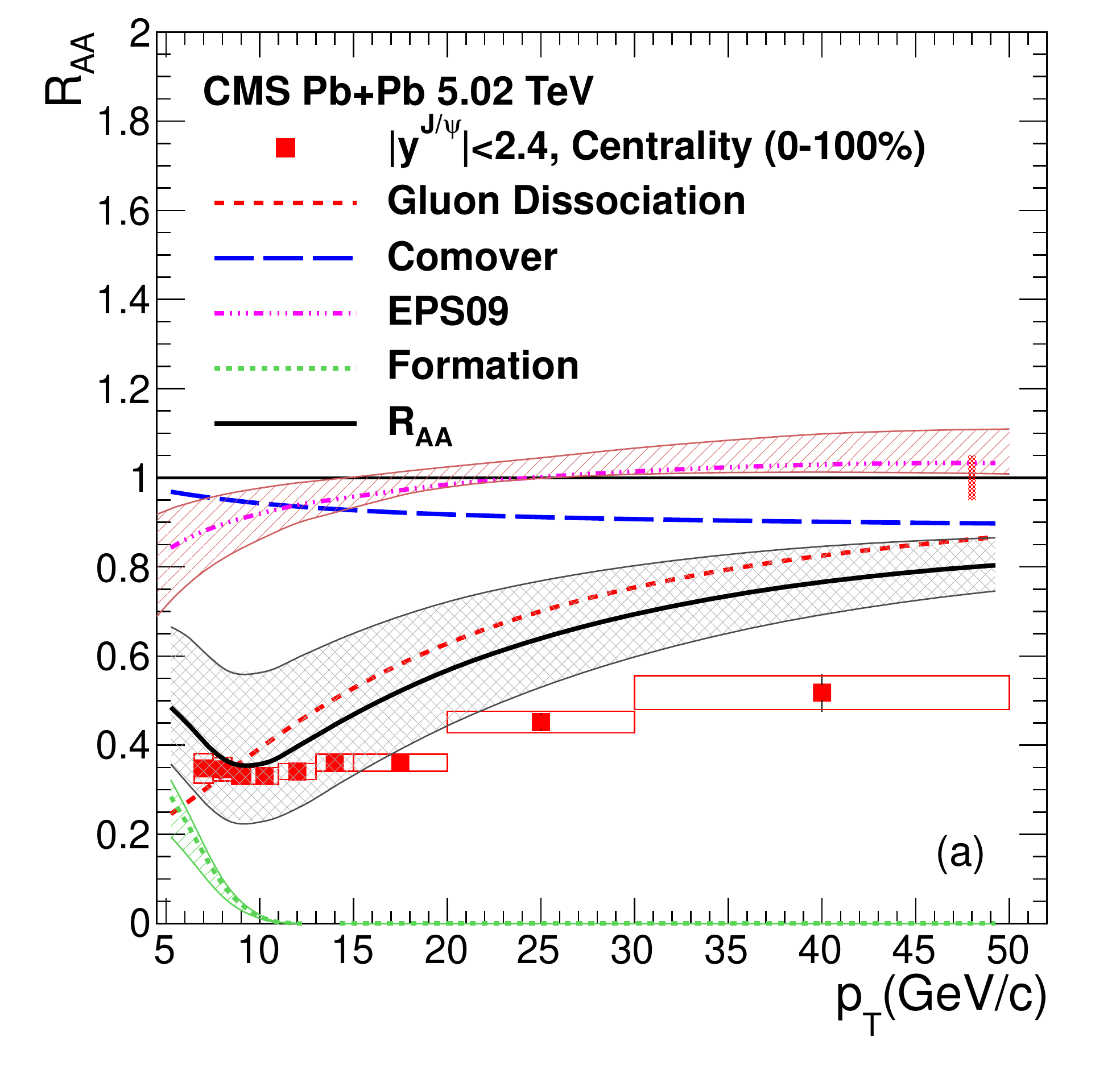}}
{\includegraphics[width=0.49\textwidth]{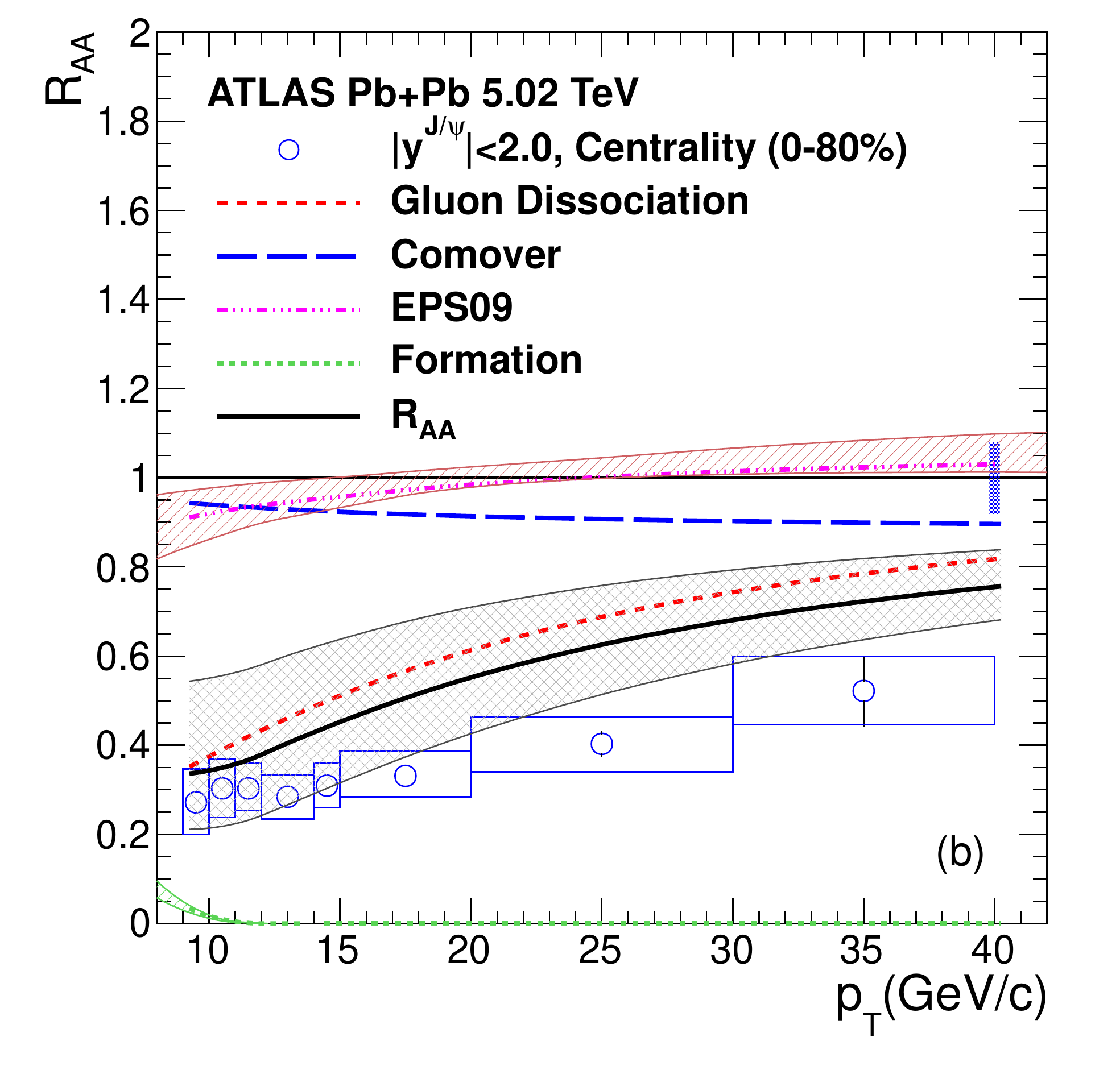}}
\end{minipage}%
\ \\
\centering
\begin{minipage}{0.5\linewidth}
\centering
{\includegraphics[width=1.0\textwidth]{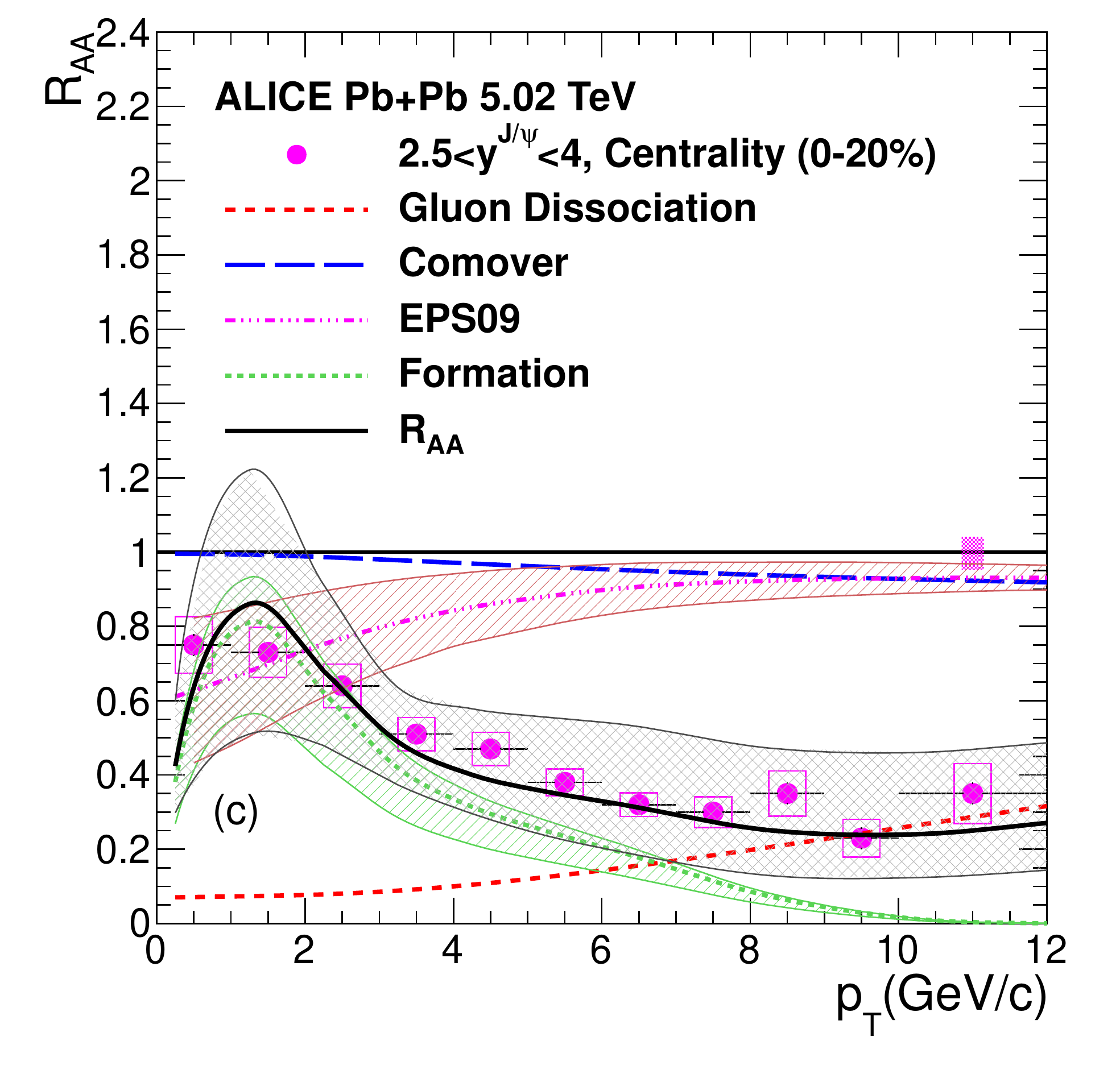}}
\end{minipage}%
\caption{(Color online)
Calculated nuclear modification factor ($R_{AA}$) as a function of $\Jpsi$ 
transverse momentum compared with (a) CMS, (b) ATLAS and (c) ALICE
measurements~\cite{Sirunyan:2017isk,ATLAS:2016qpn,Adam:2016rdg}.
The global uncertainty in $R_{AA}$ is shown as a band around the line at 1.
}
\label{fig:JPsiRaaVsPt}
\end{figure}

\begin{figure}
\begin{minipage}{1.0\linewidth}
\centering
{\includegraphics[width=0.49\textwidth]{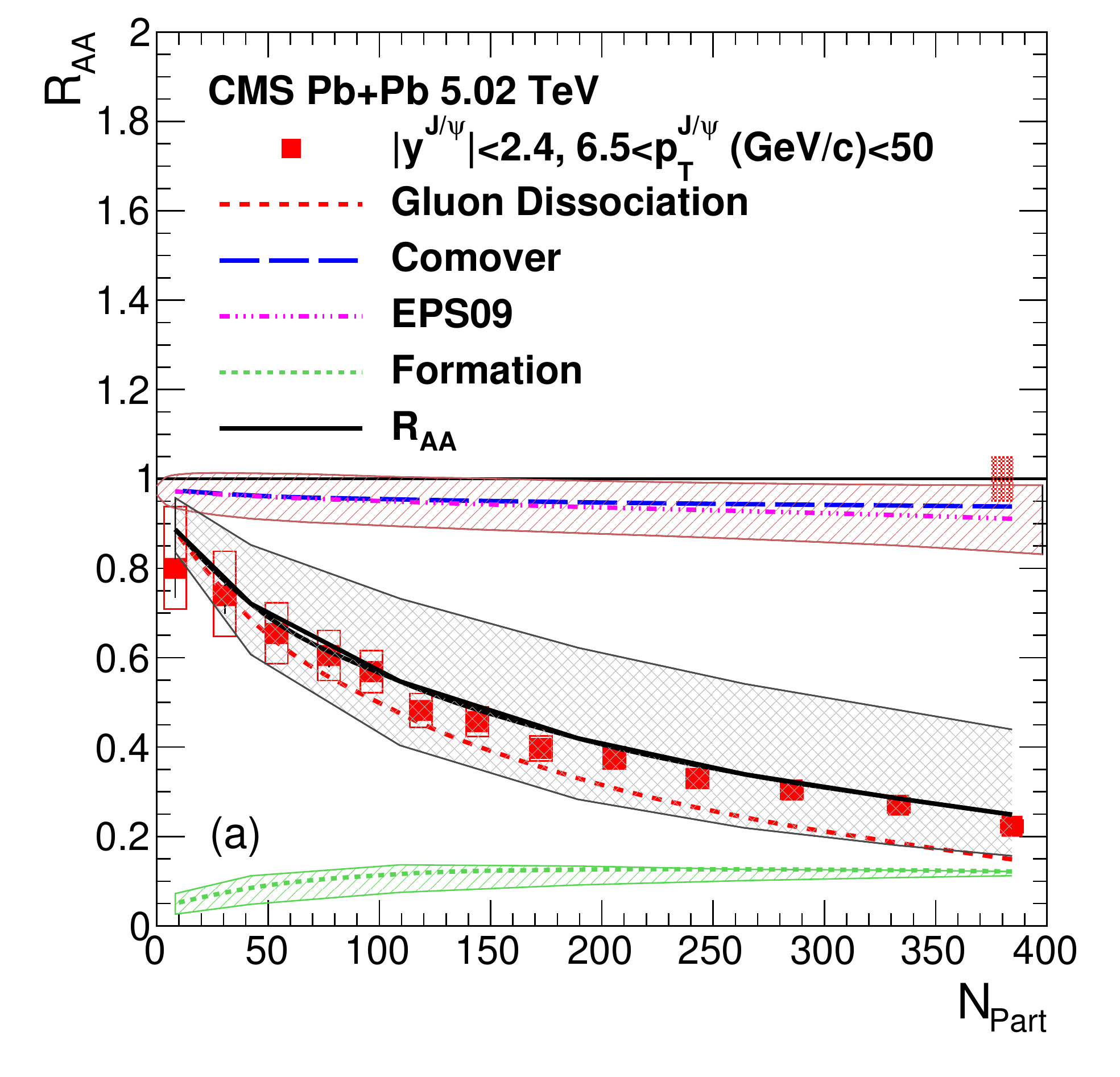}}
{\includegraphics[width=0.49\textwidth]{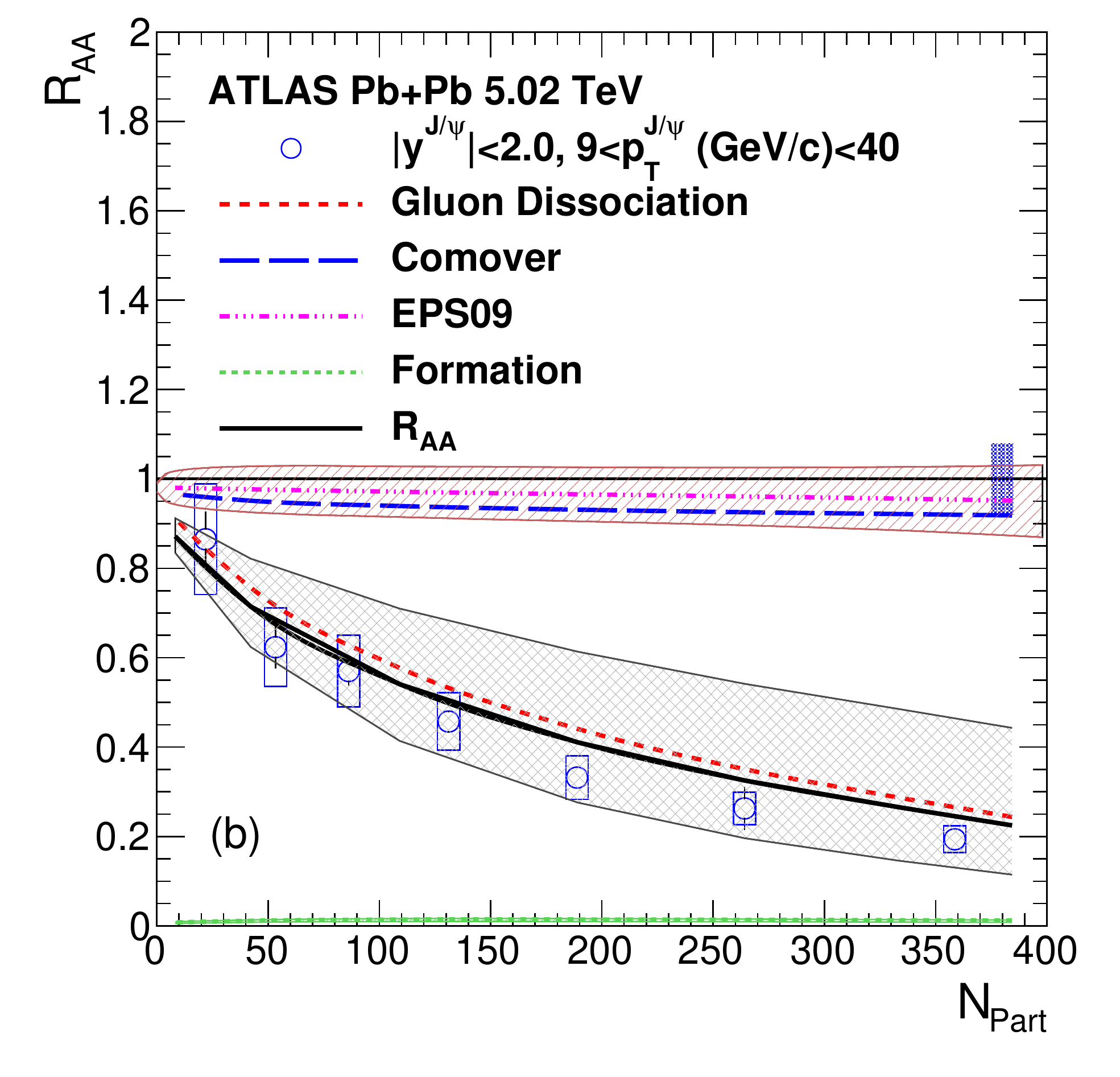}}
\end{minipage}%
\ \\
\centering
\begin{minipage}{0.5\linewidth}
\centering
{\includegraphics[width=1.0\textwidth]{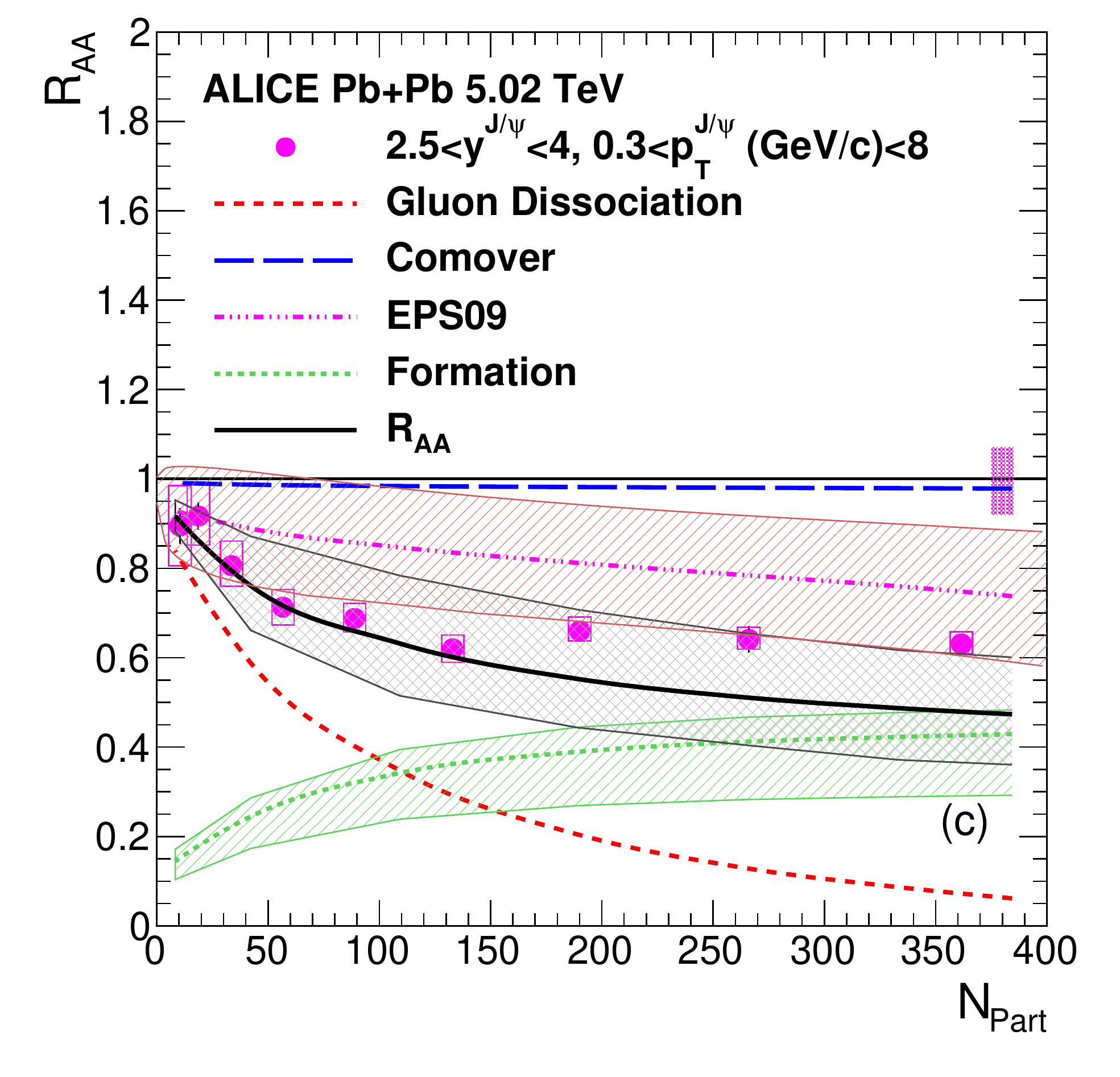}}
\end{minipage}%
\caption{(Color online) Calculated nuclear modification factor ($R_{AA}$) of $\Jpsi$ as a function
  of centrality of collisions, compared with (a) CMS, (b) ATLAS and (c) ALICE
  measurements~\cite{Sirunyan:2017isk,ATLAS:2016qpn,Adam:2016rdg}.
The global uncertainty in $R_{AA}$ is shown as a band around the line at 1.
}
\label{fig:JPsiRaaVsNPart}
\end{figure}

\begin{figure}
\includegraphics[width=0.49\textwidth]{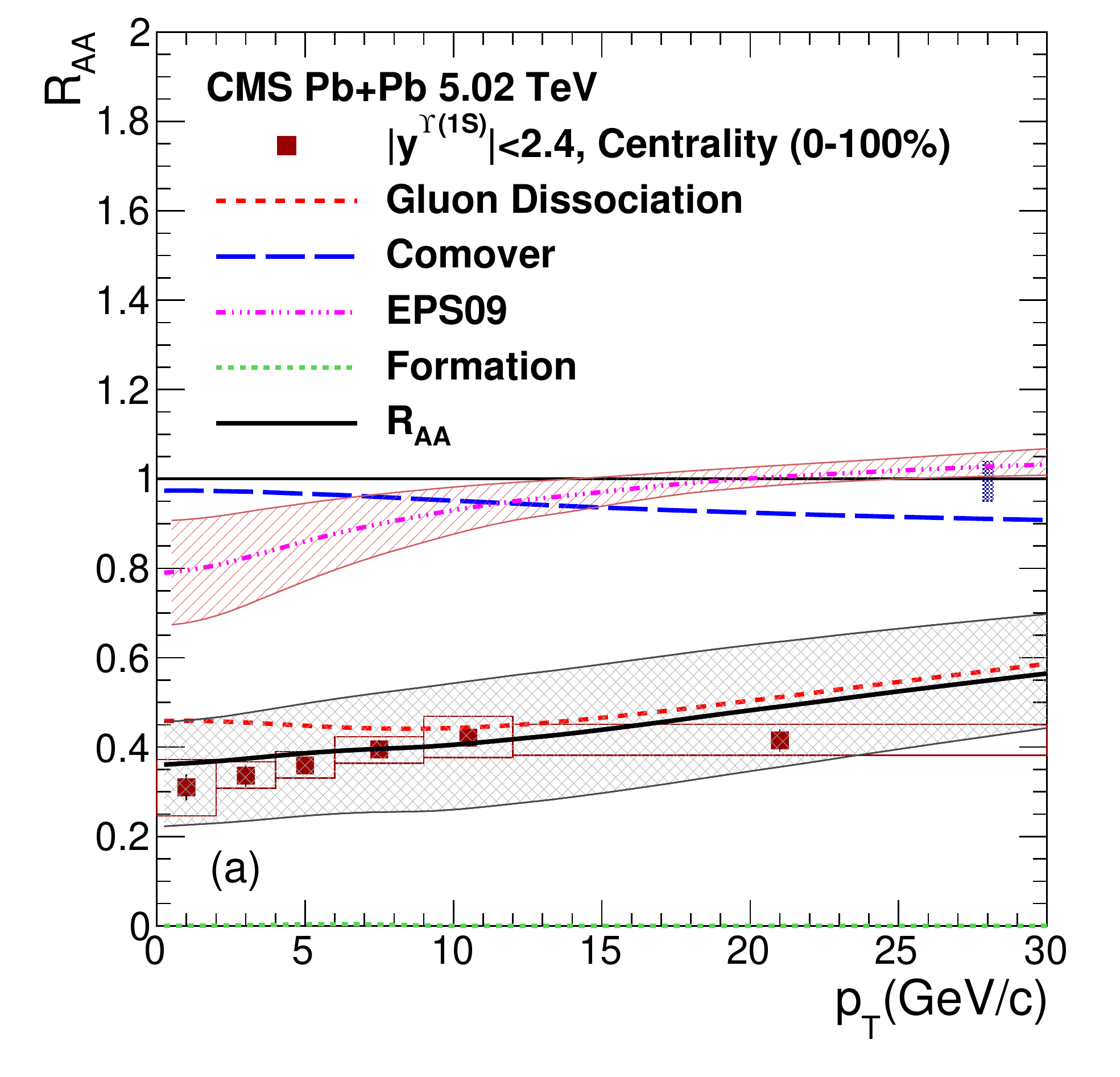}
\includegraphics[width=0.49\textwidth]{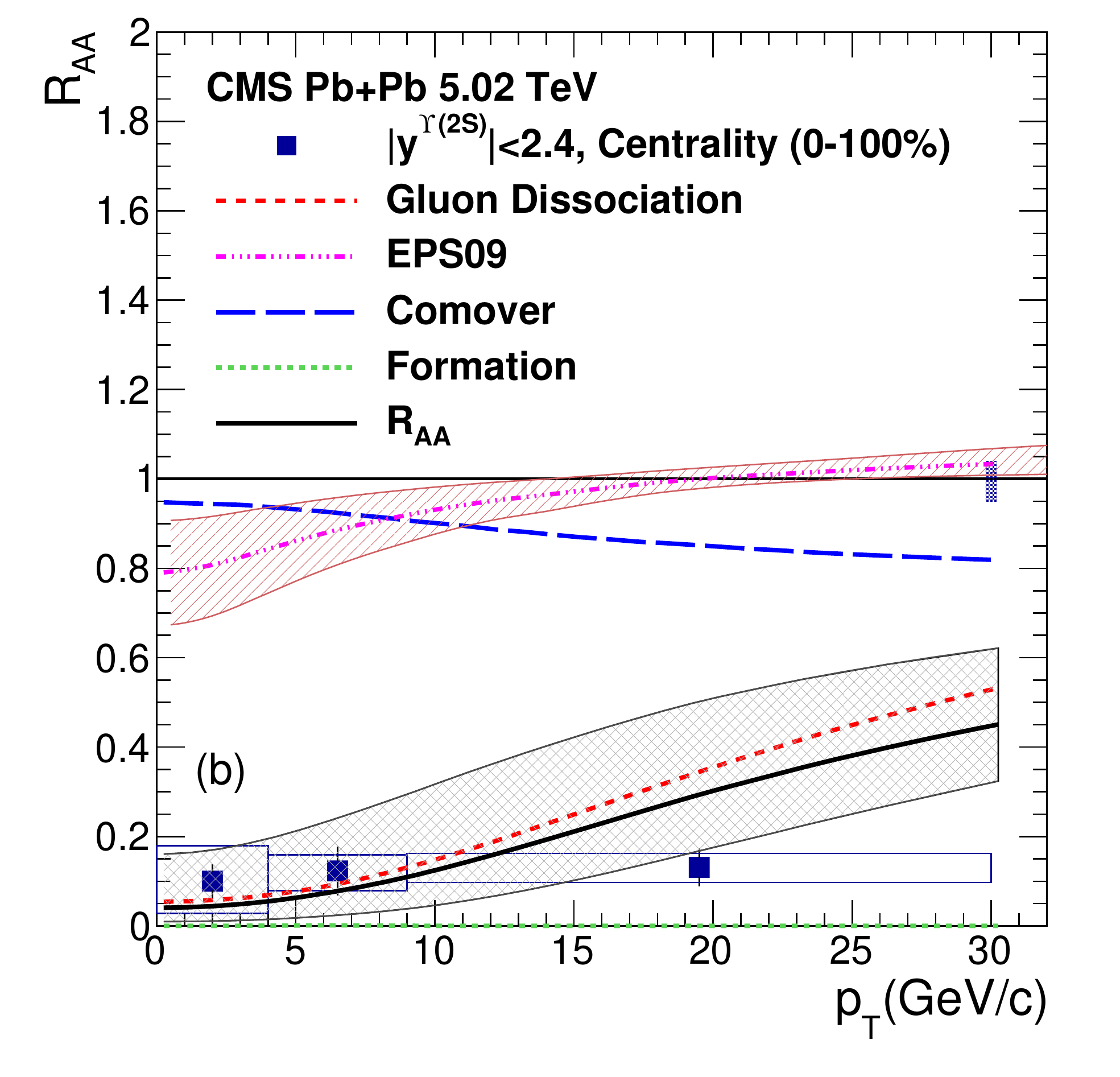}
\caption{(Color online) Calculated nuclear modification factor ($R_{AA}$) of (a) $\Upsilon$(1S) and 
  (b) $\Upsilon$(2S) as a function of $p_{T}$ 
  compared with CMS measurements~\cite{Sirunyan:2018nsz}.
The global uncertainty in $R_{AA}$ is shown as a band around the line at 1.
}
\label{fig:UpsilonRaaPtCMS}
\end{figure}

\begin{figure}
\includegraphics[width=0.49\textwidth]{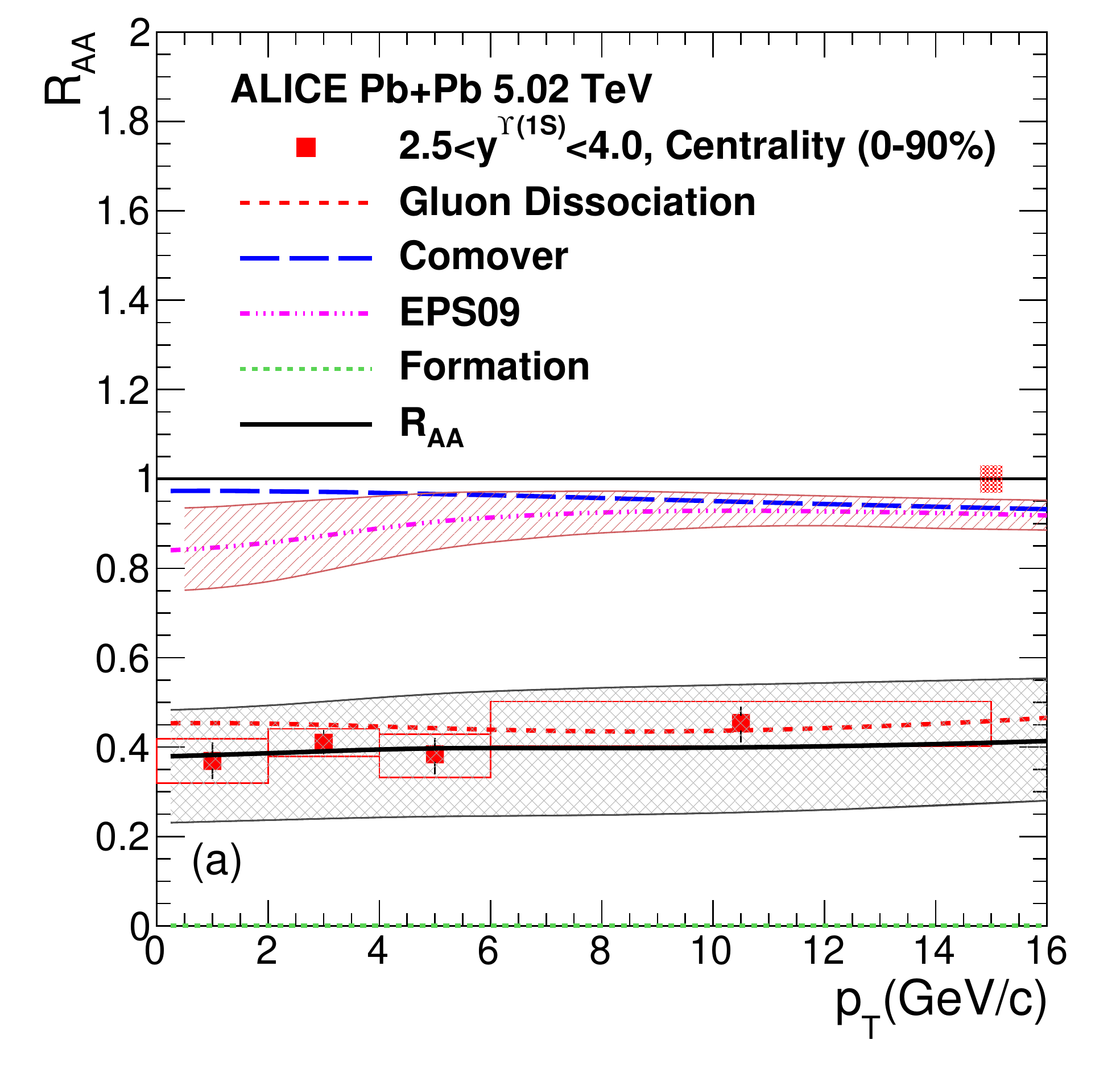}
\includegraphics[width=0.49\textwidth]{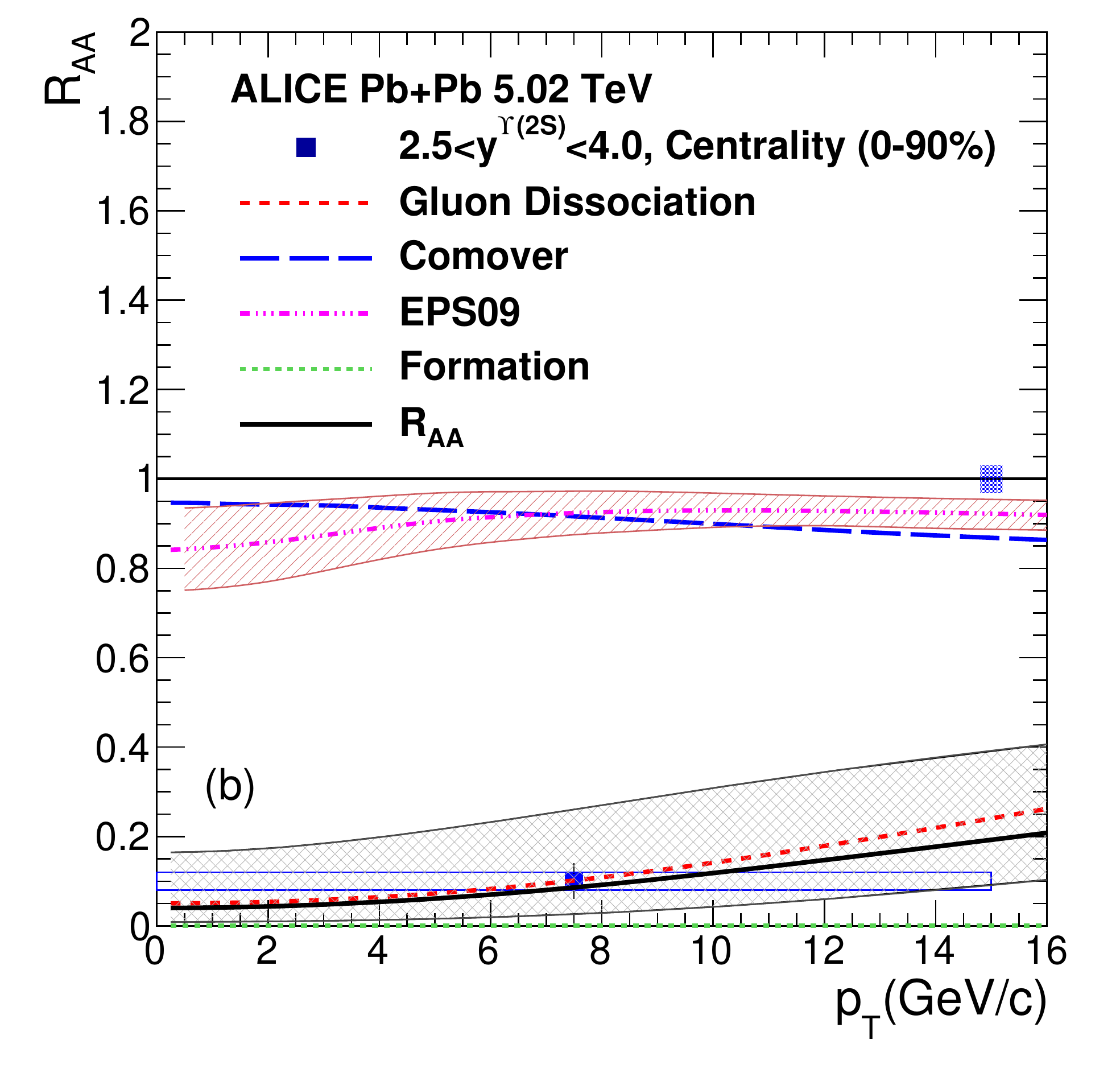}
\caption{(Color online) Calculated nuclear modification factor ($R_{AA}$) of (a) $\Upsilon$(1S) and 
  (b) $\Upsilon$(2S) as a function of $p_{T}$ in the kinematic range of ALICE detector at LHC ~\cite{ALICE:Y5TeV}.
  The global uncertainty in $R_{AA}$ is shown as a band around the line at 1.
} 
  
\label{fig:UpsilonRaaPtALICE}
\end{figure}

\begin{figure}
\includegraphics[width=0.49\textwidth]{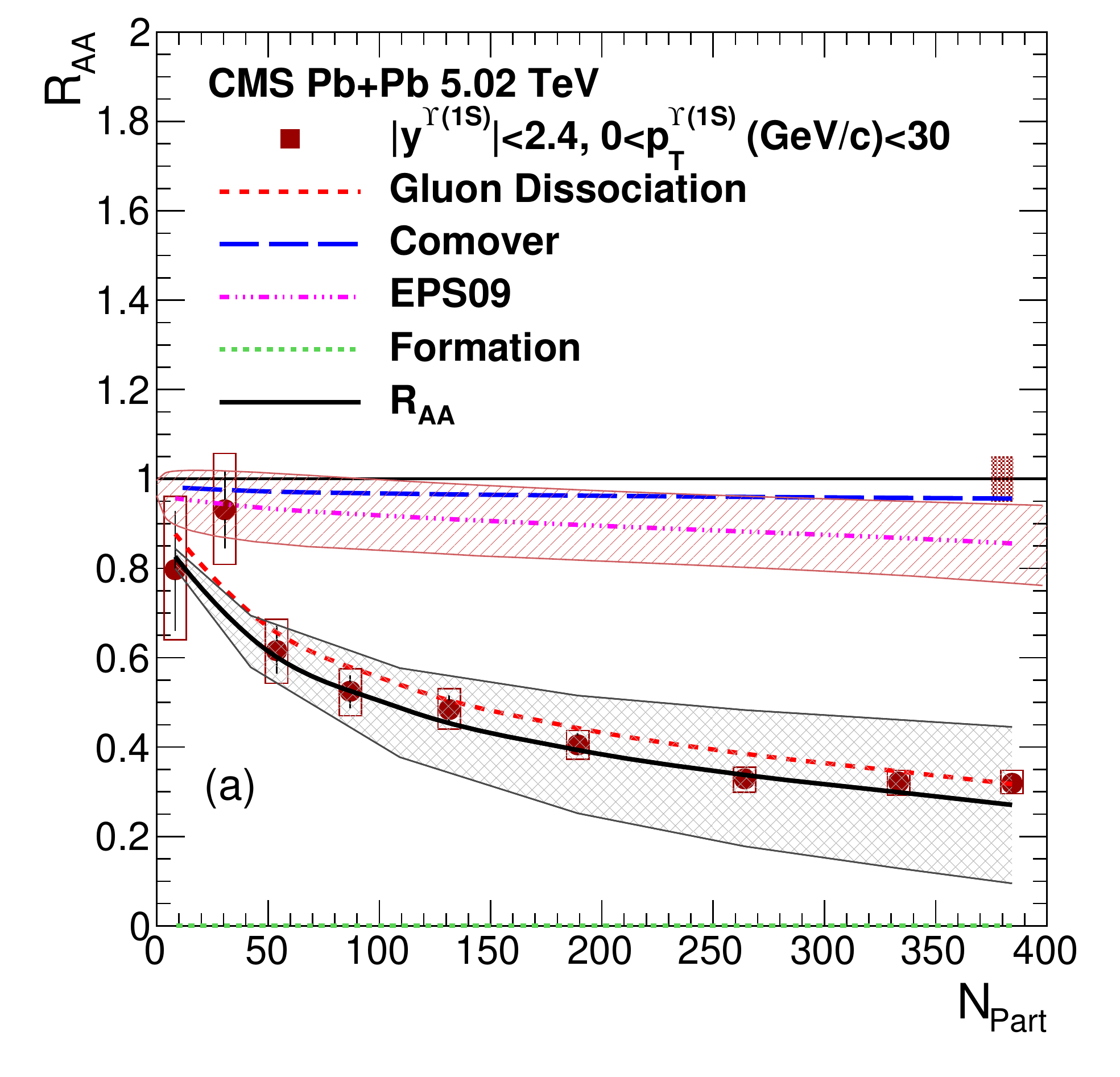}
\includegraphics[width=0.49\textwidth]{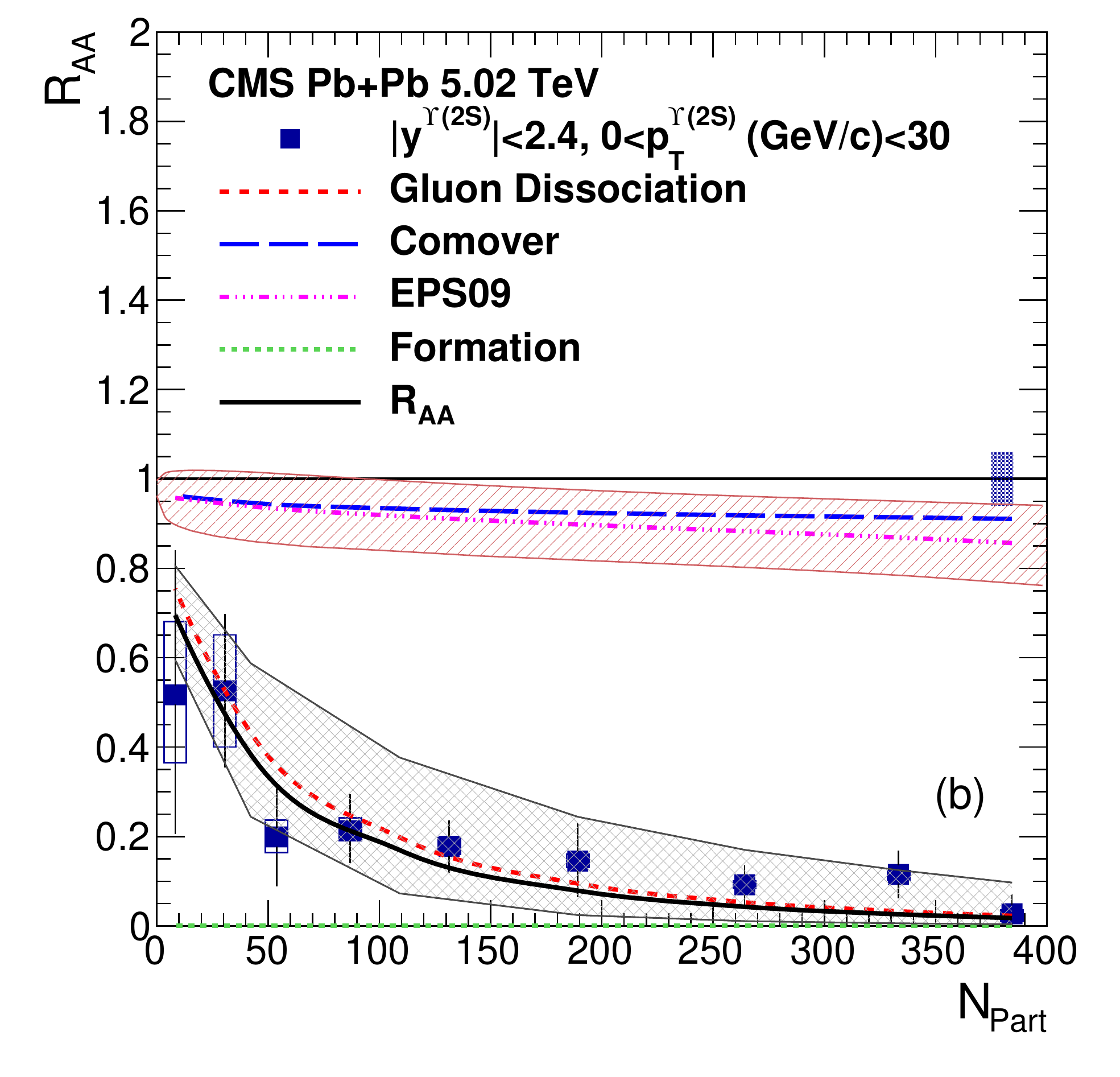}
\caption{(Color online) Calculated nuclear modification factor ($R_{AA}$) of 
  (a) $\Upsilon$(1S) and (b) $\Upsilon$(2S) as a function of centrality of the 
  collisions compared with the CMS measurements~\cite{Sirunyan:2018nsz}.
  The global uncertainty in $R_{AA}$ is shown as a band around the line at 1.
}
\label{fig:UpsilonRaaNPartCMS}
\end{figure}

Figure~\ref{fig:JPsiRaaVsPt}(a) and (b) show the estimations of different contributions
to the nuclear modification factor, $R_{AA}$, for the J/$\psi$ meson as a function of $p_T$ 
along with the mid rapidity and high $p_T$ measurements from
CMS~\cite{Sirunyan:2017isk} and ATLAS~\cite{ATLAS:2016qpn} experiments respectively.
Figure~\ref{fig:JPsiRaaVsPt}(c) shows the same for the low $p_T$ and forward
rapidity compared with the measurement by ALICE experiment~\cite{Adam:2016rdg}.
At low $p_T$, regeneration of J/$\psi$ gives dominant
contribution which overcomes the strong suppression by gluon dissociation.
This looks to be the reason for the increase of the $R_{AA}$ of J/$\psi$  around $p_T\approx$ 2 GeV/c. 
The suppression due to gluon dissociation is substantial at low $p_T$ and reduces as
we move to higher $p_T$. At low and intermediate $p_T$, both regeneration and dissociation, due
to the presence of QGP are effective.  The high $p_T$
suppression ($p_T > 10$  GeV/$c$) of $\Jpsi$ measured by CMS is greater than the
suppression caused by gluon dissociation in the QGP. The suppression measured
by CMS and ATLAS is at high $p_T$ values greater than the heavy quark mass,
and thus here the energy loss from initial partonic scatterings might play a
crucial role as it does for open heavy flavour.
{\color{black} 
The feed-down contributions for the J/$\psi$ are not very
large. At the LHC energies, around 80\% J/$\psi$ are from the directly produced hard scattering~\cite{Lansberg:2019adr}.
Moreover, the gluon dissociation cross section for the excited charmonia states are not reliable. Thus we chose
not to consider the feed-down from the higher states for the J/$\psi$.
}
The band around the total $R_{AA}$ includes all the uncertainties in the dissociation and regeneration
processes as discussed in the next paragraph. The band around CNM effects is shown
separately and is not included in the total uncertainty band for a better display.
The values of gluon-quarkonia cross section ($\sigma_{D}$) and the initial temperature $T_{0}$
can have uncertainties and will affect the results.
The value of $\sigma_D$ is varied by $\pm$50\% around the calculated value to obtain
the uncertainty in $R_{AA}$.  The initial temperature is calculated using measured
charged particle density and nominal value of $\tau_0 =$ 0.3 fm/$c$. The value of $\tau_0$
is varied  in the range 0.1 $<\,\tau_0\,<$ 0.6 fm/$c$ which corresponds to the
variation in the initial temperature from +45 \% to - 20 \%.
The uncertainty in the charm pair cross section is also considered as a source while
obtaining the contribution due to regeneration.
The total uncertainty is obtained by adding in quadrature all individual uncertainties.

The calculation of $R_{AA}$ of J/$\psi$ is also made as a function of collision
centrality (system size). 
Figure~\ref{fig:JPsiRaaVsNPart} shows calculations of different contributions to the J/$\psi$ 
nuclear modification factor as a function of system size, along with the measurements
from CMS in (a), ATLAS in (b) and ALICE in (c)~\cite{Sirunyan:2017isk,ATLAS:2016qpn,Adam:2016rdg}.
The figure shows that the suppression of J/$\psi$ due to
QGP increases when the system size grows. The contribution from regeneration process
is minimum in the high $p_T$ range ($p_T~>9$ GeV/c) for ATLAS case
as shown in Fig.~\ref{fig:JPsiRaaVsNPart}(b) and maximum in the low $p_T$ range
for the ALICE case shown in Fig.~\ref{fig:JPsiRaaVsNPart}(c). For low $p_T$ ALICE measurement,
the nuclear modification factor is almost flat from mid to central collisions because
the regeneration compensates the gluon dissociation, a trend which is well-reproduced
by our calculations.
The CMS centrality dependence of the $R_{AA}$ of $\Jpsi$ given in
Fig.~\ref{fig:JPsiRaaVsNPart}(a) is reasonably well described by the model since 
the contribution of the CMS data comes from $\Jpsi$ with 6.5 $<\,p_T<\,$10 GeV/$c$
where the suppression due to gluon dissociation dominates.
For the case of ATLAS data, the model reproduces the shape of centrality dependence
of the $R_{AA}$ of $\Jpsi$ observed in the data.

\begin{figure}
\includegraphics[width=0.49\textwidth]{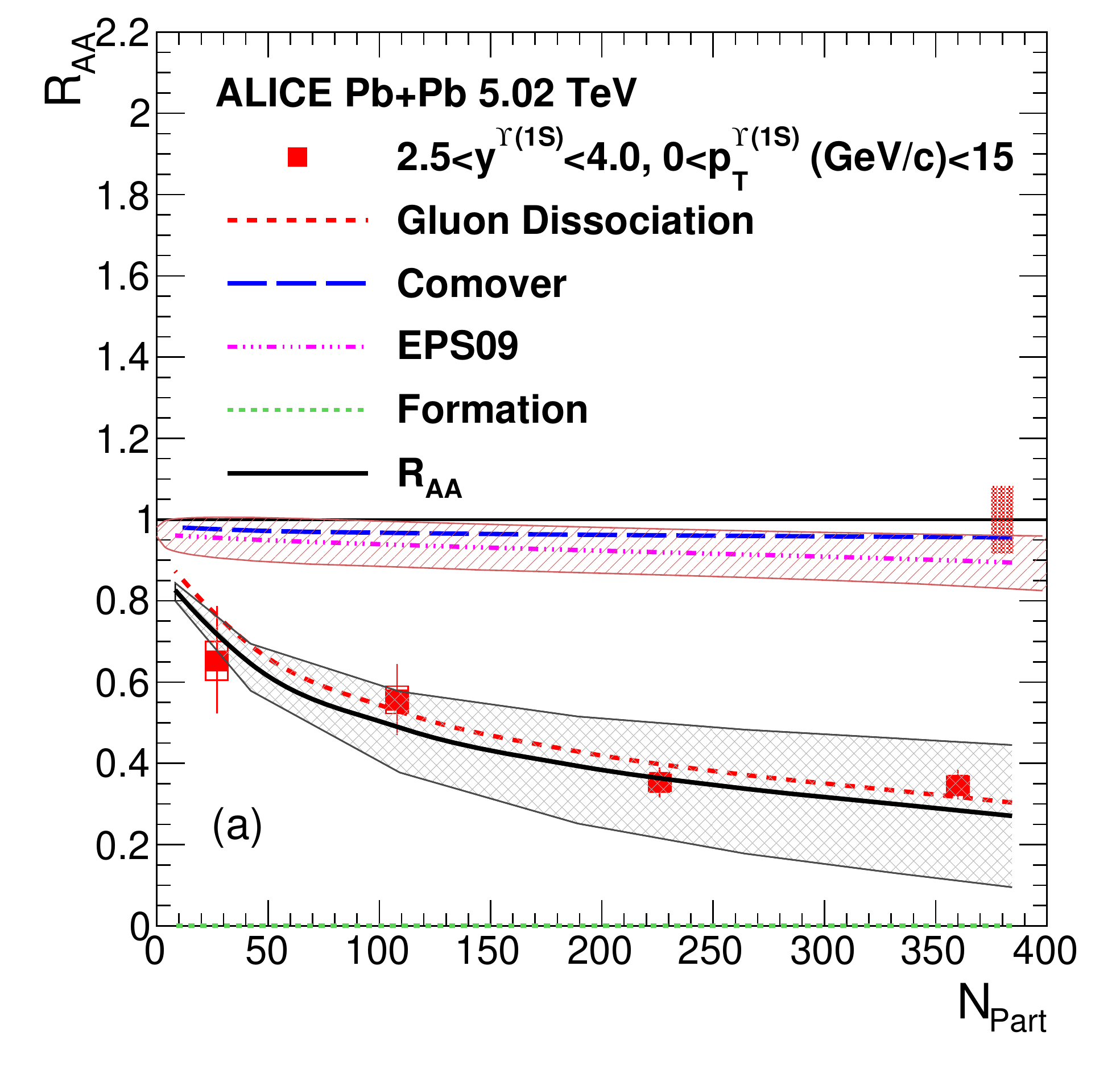}
\includegraphics[width=0.49\textwidth]{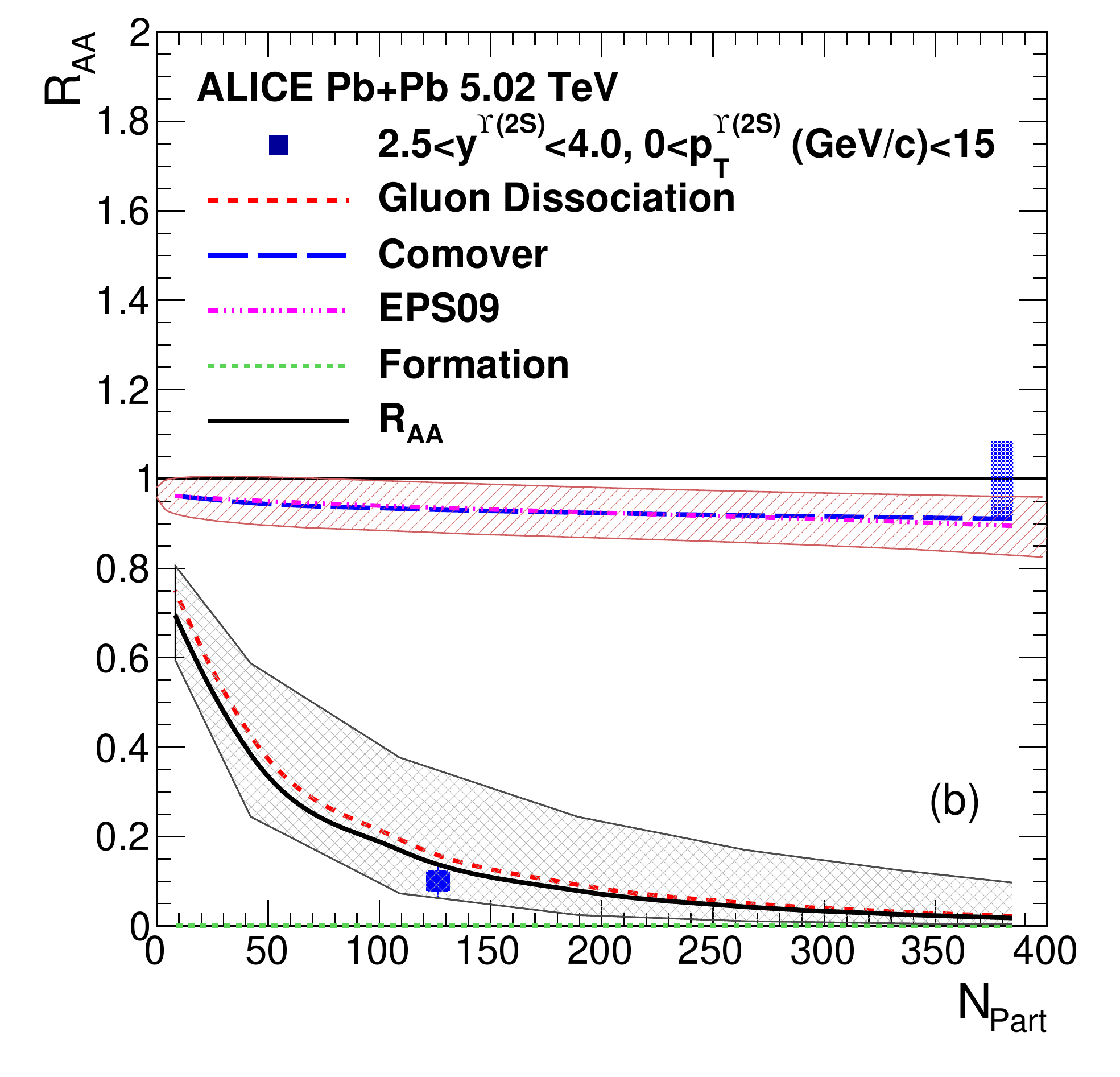}
\caption{(Color online) Calculated nuclear modification factor ($R_{AA}$) of 
  (a) $\Upsilon$(1S) and (b) $\Upsilon$(2S) as a function of centrality of 
  the collisions compared with the ALICE measurement~\cite{ALICE:Y5TeV}.
The global uncertainty in $R_{AA}$ is shown as a band around the line at 1.
}
\label{fig:UpsilonRaaNPartALICE}
\end{figure}

Figure~\ref{fig:UpsilonRaaPtCMS}(a) and (b) show the calculations of contributions to
the nuclear modification factor, $R_{AA}$, for the $\Upsilon$(1S) and $\Upsilon$(2S)
respectively as a function of $p_T$ compared with the mid rapidity measurements from
CMS~\cite{Sirunyan:2018nsz}.  
The gluon dissociation mechanism combined with the pion dissociation and shadowing
corrections gives good description of data in mid $p_{T}$ range ($p_{T}\approx$ 5-10 GeV/c)
for both $\Upsilon$(1S) and $\Upsilon$(2S).
The contribution from the regenerated $\Upsilon$s is negligible even at LHC energies.
Our calculations under-predict the suppression observed at the highest measured
$p_{T}$ for $\Upsilon$(1S) and $\Upsilon$(2S) which is similar for the case
of J/$\psi$.
The states $\Upsilon$(1S) and $\Upsilon$(2S) also have
feed-down contributions from decays of higher b$\bar{\rm b}$ bound states.
The nuclear modification factor, $R_{AA}$ is obtained taking into account the feed-down corrections as follows
  \begin{equation}
    R_{AA}^{\Upsilon(3S)} = R_{AA}^{\Upsilon(3S)}\\ 
  \end{equation}
  \begin{equation}
    R_{AA}^{\Upsilon(2S)} = f_1 R_{AA}^{\Upsilon(2S)} +  f_2 R_{AA}^{\Upsilon(3S)} \\ 
  \end{equation}
   \begin{equation}
    R_{AA}^{\Upsilon(1S)} = g_1 R_{AA}^{\Upsilon(1S)} +  g_2 R_{AA}^{\chi_b(1P)} + g_3 R_{AA}^{\Upsilon(2S)} + g_4 R_{AA}^{\Upsilon(3S)}\\ 
  \end{equation}
The factors $f$’s and $g$’s are obtained from CDF measurement~\cite{Affolder:1999wm}.
The values of $g_1$, $g_2$, $g_3$ and $g_4$ are 0.509, 0.27, 0.107
and 0.113 respectively. Here $g_4$ is assumed to be the combined fraction of 
$\Upsilon$(3S) and $\chi_b$(2P).
The values of $f_1$ and $f_2$ are taken as 0.50~\cite{Strickland:2011aa}.

Figure~\ref{fig:UpsilonRaaPtALICE}(a) and (b) show the model 
prediction of the nuclear modification factor, $R_{AA}$, for the $\Upsilon$(1S)
and $\Upsilon$(2S) respectively as a function of $p_T$ in the kinematic range
covered by ALICE detector. The ALICE data~\cite{ALICE:Y5TeV} is well described by our model.

Figure~\ref{fig:UpsilonRaaNPartCMS}(a) depicts the calculated 
centrality dependence of the $\Upsilon$(1S) nuclear
modification factor, along with the midrapidity data from CMS~\cite{Sirunyan:2018nsz}.
Our calculations combined with the pion dissociation and shadowing corrections 
gives very good description of the measured data. Figure~\ref{fig:UpsilonRaaNPartCMS}(b)
shows the same for the $\Upsilon$(2S) along with the midrapidity
CMS measurements. The suppression of the excited $\Upsilon$(2S) states 
is also well described by our model. As stated earlier, the effect of regeneration is
negligible for $\Upsilon$ states. 

Figure~\ref{fig:UpsilonRaaNPartALICE}(a) shows the forward rapidity ALICE
measurement of the $\Upsilon$(1S) nuclear modification factor~\cite{ALICE:Y5TeV}
along with our calculations. The suppression due to thermal gluon dissociation 
describes the measured data after including the comover and shadowing corrections.
Figure~\ref{fig:UpsilonRaaNPartALICE}(b) shows the calculations for the
$\Upsilon$(2S) nuclear modification factor in ALICE detector kinematic range.
The suppression due to thermal gluon dissociation describes the
ALICE measurements after including the comover and shadowing corrections.

\section{Summary}
 We presented detailed calculations of the $\Jpsi$ and $\Upsilon$ 
production and the modifications of their yields in PbPb collisions at $\sNN =$ 5.02 TeV.
A kinetic model is employed which incorporates quarkonia suppression inside QGP, suppression 
due to hadronic comovers and regeneration from heavy quark pairs.
The behaviour of the dissociation and formation rates are studied as a function of
transverse momentum and medium temperature. 
The nuclear modification factors for both $\Jpsi$ and $\Upsilon$ are obtained 
as a function of system size and transverse momentum and have been compared to the measurements
in PbPb collisions at $\sNN =$ 5.02 TeV.
It is found that  regeneration of $\Jpsi$ is the dominant process at low $p_T$. As a result
the $\Jpsi$ production is found to be enhanced in the ALICE low $p_T$ data.
In the same $p_T$ range gluon dissociation is also substantial however it becomes small
as we move to high $p_T$. 
Both  these processes affect the yields of quarkonia in a QGP medium  at low and
intermediate $p_T$.
 The high $\pT$ suppression ($p_T > 10$  GeV/$c$) of $\Jpsi$ measured by CMS and
ATLAS is more than the suppression expected due to gluon dissociation in QGP.
The energy loss from initial partonic scatterings might play a crucial role in this region
as it does for open heavy flavour.
As the system size grows  $\Jpsi$'s are increasingly suppressed.  The nuclear 
modification factor at low $p_T$  (ALICE case) as a function of centrality remains
flat since the increased suppression is compensated by regenerated  $\Jpsi$'s
as the system size grows. 
 We could reproduce the centrality dependence of $R_{AA}$ for high $p_T$ $\Jpsi$'s reasonably
 well. The $\pT$ and centrality dependence of suppression of $\Upsilon$ states are well reproduced
 by the model.

\section{Acknowledgement}
Authors thank Board of Research in Nuclear Sciences (BRNS) for support. 
AB thanks Alexander von Humboldt (AvH) foundation and Federal Ministry 
of Education and Research (Gernamy) for support through Research Group 
Linkage Programme. 


\section*{References}

\end{document}